\documentclass[useAMS,usenatbib]{mn2e}

\usepackage{color}
\usepackage{graphics}
\usepackage{psfig,subfigure}
 
\title[Mass Accretion onto T Tauri Stars]
      {Mass Accretion onto T Tauri Stars}
\author[S. G. Gregory, M. Jardine, I. Simpson and J.- F. Donati]
{S. G. Gregory$^{1}$\thanks{E-mail: sg64@st-andrews.ac.uk},
 M. Jardine$^{1}$, I. Simpson$^{1}$ and J.- F. Donati$^{2}$ \\
$^{1}$School of Physics and Astronomy, University of St Andrews, North 
Haugh, St Andrews, Fife, KY16 9SS, U. K.\\
$^{2}$Laboratoire d'Astrophysique, Observatoire Midi-Pyr\'en\'ees, 
      14 Av. E. Belin, F-31400 Toulouse, France}

\begin{document}

\date{}

\pagerange{\pageref{firstpage}--\pageref{lastpage}} \pubyear{2006}

\maketitle

\label{firstpage}

\begin{abstract}
It is now accepted that accretion onto classical T Tauri stars
is controlled by the stellar magnetosphere, yet to date most accretion 
models have assumed that their magnetic fields are dipolar.  By considering 
a simple steady state accretion model with both dipolar and complex 
magnetic fields we find a correlation between mass accretion rate and stellar 
mass of the form $\dot{M} \propto M_{\ast}^{\alpha}$, with our results consistent
within observed scatter.  For any particular stellar mass there can be several 
orders of magnitude difference in the mass accretion rate, with accretion 
filling factors of a few percent.  We demonstrate that the field 
geometry has a significant effect in controlling the location and 
distribution of hot spots, formed on the stellar surface from the high 
velocity impact of accreting material.  We find that hot spots are often at 
mid to low latitudes, in contrast to what is expected for accretion to dipolar 
fields, and that particularly for higher mass stars, the accretion flow is 
predominantly carried by open field lines.     
\end{abstract}


\section{Introduction}
Classical T Tauri stars (CTTSs) are young, low mass, pre-main sequence 
stars that are actively accreting from a surrounding disc which is the eventual 
birth-place of planets. \citet{uch84} suggested that the magnetic 
field of a CTTS disrupts the inner disc.  In the early 1990s several 
magnetospheric accretion models were developed (\citealt{kon91};
\citealt{col93}; \citealt{shu94}) where material is lifted from
the disc plane and is channelled along dipolar magnetic field lines
onto the star,  terminating in a shock at the photosphere.    
In an idealised model of a CTTS's magnetic field there are closed field 
lines close to the star that contain the X-ray emitting 
corona, whilst at larger radii, there are closed field lines which 
thread the circumstellar disc.  It is along this latter set of field 
lines that accretion may proceed.  There are also regions of open 
field which carry outflows in the form of a wind, and in some cases, 
as large collimated bipolar jets.  

Magnetospheric accretion models assume that CTTSs possess magnetic 
fields that are strong enough to disrupt the disc at a distance of 
a few stellar radii.  Such strong fields have been detected in a 
number of systems using a variety of techniques.  Average surface fields 
of 1-3kG have been detected most successfully by exploiting the Zeeman effect, 
both through Zeeman broadening (e.g. \citealt*{joh99b}) and from the circular 
polarisation of lines which are sensitive to the presence of a magnetic 
field (e.g. \citealt{joh99a}; \citealt{sym05}; \citealt*{dao06}).  Field detections have also 
been made from the increase in line equivalent width (\citealt*{bas92}; 
\citealt{gue99}) and also from electron cyclotron maser emission, a coherent 
emission process from mildly relativistic electrons trapped inside flux tubes 
close to the star \citep{smi03}.  The mean magnetic field strengths detected 
so far appear to be roughly constant across all stars \citep{val04}.    

Traditionally magnetospheric accretion models have assumed the CTTSs 
have dipolar magnetic fields.  Dipole fields (or inclined dipole fields) have 
been successively used to explain some of the observations of CTTSs (e.g. the 
photopolarimetric variability of AA Tau, \citealt{osu05}), but fail to 
account for others.  \citet{val04} present magnetic field measurements
for a number of stars, and despite detecting strong average surface 
fields from Zeeman broadening, often measurements of the longitudinal 
(line-of-sight) field component (obtained from photospheric lines) 
are consistent with no net circular polarisation.  This can be interpreted 
as there being many regions of opposite polarity on the stellar surface, giving 
rise to oppositely polarised signals which cancel each other out giving 
a net polarisation signal of zero.  This suggests that CTTSs have magnetic 
fields which are highly complex, particularly close to the stellar surface; 
however, as \citet{val04} point out, as the higher order multi-pole field 
components will drop off quickly with distance from the star, the dipole 
component may still remain dominant at the inner edge of the disc.  Also 
their measurements of the circular polarisation of the HeI 5876{\AA} emission 
line (believed to form in the base of accretion columns) are well fitted by a 
simple model of a single magnetic spot on the surface of the star, suggesting that 
the accreting field may be well ordered, despite the surface field being
complex.  

The fractional surface area of a CTTS which is covered in hot spots, the 
accretion filling factor $f_{acc}$, is inferred from observations to be small; typically 
of order one percent (\citealt{muz03}; \citealt{cal04}; \citealt{val04}; 
\citealt{sym05}; \citealt{muz05}).  Dipolar magnetic field models predict
accretion filling factors which are too large.  This, combined with the 
polarisation results, led \citet{joh02} to generalise the Shu X-wind
model \citep{shu94} to include multipolar, rather than dipolar, magnetic
fields.  With the assumption that the average surface field strength 
does not vary much from star to star the generalised Shu X-wind model
predicts a correlation between the stellar and accretion parameters      
of the form $R_{\ast}^2f_{acc} \propto (M_{\ast}\dot{M}P_{rot})^{1/2}$, a prediction
that matches observations reasonably well.

In this paper we present a model of the accretion process using both 
dipolar and complex magnetic fields.  We apply our model to a large sample 
of pre-main sequence stars obtained from the Chandra Orion Ultradeep 
Project (COUP; \citealt{get05}), in order to test if our model can reproduce the 
observed correlation between mass accretion rate and stellar mass.  An increase 
in  $\dot{M}$ with $M_{\ast}$ was originally noted by \citet{reb00} and 
subsequently by  \citet{whi01} and \citet{reb02}.  The correlation was then 
found to extend to very low mass objects and accreting brown dwarfs by 
\citet{whi03} and \citet{muz03}, and to the higher mass, intermediate mass 
T Tauri stars, by \citet{cal04}.  Further low mass data has recently been added 
by \citet{nat04}, \citet*{moh05} and by \citet{muz05} who obtain a correlation of 
the form $\dot{M} \propto M_{\ast}^{2.1}$, with as much as three orders of 
magnitude scatter in the measured mass accretion rate at any particular stellar 
mass.  However, \citet{cal04} point out that due to a bias against the detection 
of higher mass stars with lower mass accretion rates, the power may be less 
than 2.1.  Further data for accreting stars in the $\rho-$Ophiuchus star forming region 
has recently been added by \citet*{nat06}.   

The physical origin of the correlation between $M_{\ast}$ and $\dot{M}$, 
and the large scatter in measured $\dot{M}$ values, is not clear; however several 
ideas have been put forward.  First, increased X-ray emission in higher mass 
T Tauri stars (\citealt{pre05}; \citealt{jar06}) may cause an increase in disc 
ionisation, leading to a more efficient magnetorotational instability and 
therefore a higher mass accretion rate \citep{cal04}.  Second, \citet{pad05} 
argue that the correlation $\dot{M} \propto M_{\ast}^{2}$ arises from 
Bondi-Hoyle accretion, with the star-disc system gathering mass as it moves 
through the parent cloud.  In their model the observed scatter in $\dot{M}$ 
arises from variations in stellar velocities, gas densities and sound speeds.  
\citet{moh05} provide a detailed discussion of both of these suggestions.
Third, \citet{ale06} suggest that the correlation may arise from variations
in the disc initial conditions combined with the resulting viscous 
evolution of the disc.  In their model they assume that the initial disc
mass scales linearly with the stellar mass, $M_d \propto M_{\ast}$, which, upon making this
assumption, eventually leads them to the conclusion that brown dwarfs 
(the lowest mass accretors) should have discs which are larger than higher mass 
accretors.  However, if it is the case that the initial disc mass increases more 
steeply with stellar mass, $M_d \propto M_{\ast}^2$, then the stellar mass - accretion 
rate correlation can be reproduced with smaller brown dwarf discs of low mass (of order 
one Jupiter mass).  Thus the \citet{ale06} suggestion, if correct, will soon
be directly verifiable by observations.  Fourth, \citet{nat06} suggest that the
large scatter in the correlation between $\dot{M}$ and $M_{\ast}$ may arise 
from the influence of close companion stars, or by time variable accretion.
It should however be noted that \citet{cla06} take a more conservative view by 
demonstrating that a steep correlation between $\dot{M}$ and $M_{\ast}$ may arise
as a consequence of detection/selection limitations, and as such 
$\dot{M} \propto M_{\ast}^{2}$ is perhaps not a true representation of the correlation 
between $\dot{M}$ and $M_{\ast}$. 

In \S2 we describe how magnetic fields are extrapolated from observed surface magnetograms. 
In \S3 we consider accretion onto an aligned, and then a tilted dipole field, to develop a simple 
steady state accretion model and to investigate how tilting the field affects 
the mass accretion rate.  In \S4 these ideas are extended by considering magnetic fields with 
a realistic degree of complexity and we apply our accretion model to study the correlation 
between mass accretion rate and stellar mass, whilst \S5 contains our conclusions.

%-------------------------------------------------------------------------------------

\section{Realistic magnetic fields}
From Zeeman-Doppler images it is possible to extrapolate stellar magnetic 
fields by assuming that the field is potential.  At the moment we 
do not have the necessary observations of CTTSs, but we do have for the solar like 
stars LQ Hya and AB Dor (\citealt{don97a}; \citealt{don97b}; \citealt{don99a}; 
\citealt{don99b}; \citealt{don03}), which have  
different field topologies (\citealt*{jar02a}; \citealt{hus02}; \citealt{mci03}; 
\citealt{mci04}). Using their field structures as an example we can adjust the 
stellar parameters (mass, radius and rotation period) to construct a simple model 
of a CTTS, surrounded by a thin accretion disc.

The method for extrapolating magnetic fields follows that employed by
\citet{jar02a}.  Assuming the magnetic field $\mathbf{B}$
is potential, or current-free, then $\nabla \times \mathbf{B} =
0$. This condition is satisfied by writing the field in terms of a
scalar flux function $\Psi$, such that $\mathbf{B}=-\nabla \Psi$.
Thus in order to ensure that the field is divergence-free ($\nabla
\cdot \mathbf{B}=0$), $\Psi$ must satisfy Laplace's equation, $\nabla^2
\Psi=0$; the solution of which is a linear combination of
spherical harmonics,
\begin{equation}
\Psi= \sum_{l=1}^{N} \sum_{m=-l}^{l} \left
[a_{lm}r^l+b_{lm}r^{-(l+1)} \right ] P_{lm}(\theta) e^{im\phi},
\end{equation}
where $P_{lm}$ denote the associated Legendre functions.  It then
follows that the magnetic field components at any point
$(r,\theta,\phi)$ are,
\begin{equation}
B_r  =  -\sum_{l=1}^{N}\sum_{m=-l}^{l}
               [la_{lm}r^{l-1} - (l+1)b_{lm}r^{-(l+2)}]
               P_{lm}(\theta) e^{i m \phi}
\end{equation}
\begin{equation}
B_\theta  =  -\sum_{l=1}^{N}\sum_{m=-l}^{l} 
               [a_{lm}r^{l-1} + b_{lm}r^{-(l+2)}]
               \frac{d}{d\theta}P_{lm}(\theta) e^{i m \phi}
\end{equation} 
\begin{equation} 
B_\phi  =  -\sum_{l=1}^{N}\sum_{m=-l}^{l} 
               [a_{lm}r^{l-1} + b_{lm}r^{-(l+2)}]
               \frac{P_{lm}(\theta)}{\sin \theta} ime^{i m \phi}.
\end{equation}
The coefficients $a_{lm}$ and $b_{lm}$ are determined from the
radial field at the stellar surface obtained from Zeeman-Doppler
maps and also by assuming that at some height
$R_s$ above the surface (known as the source surface) 
the field becomes radial and hence
$B_{\theta}(R_s)=0$, emulating the effect of the corona 
blowing open field lines to form a stellar wind \citep{alt69}.  
In order to extrapolate the field we used a modified version of a 
code originally developed by \citet*{van98}.

\begin{figure*}
        \def\subfigtopskip{4pt}
        \def\subfigbottomskip{4pt}
        \def\subfigcapskip{2pt}
        \centering
        \begin{tabular}{cc}
        \subfigure[]{
                        \label{lqhya_extrap}                    
                        \psfig{figure=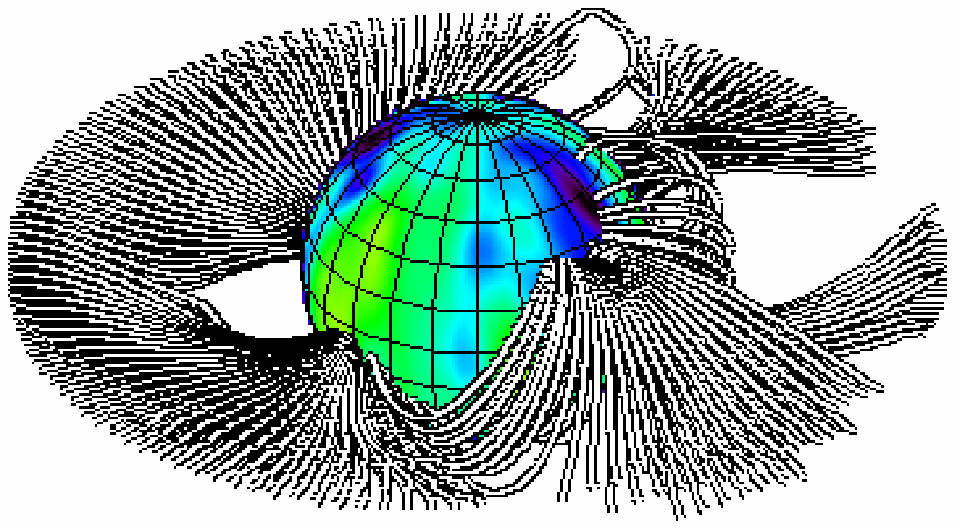,width=80mm}
                        } &
                \subfigure[]{
                        \label{abdor_extrap} 
                        \psfig{figure=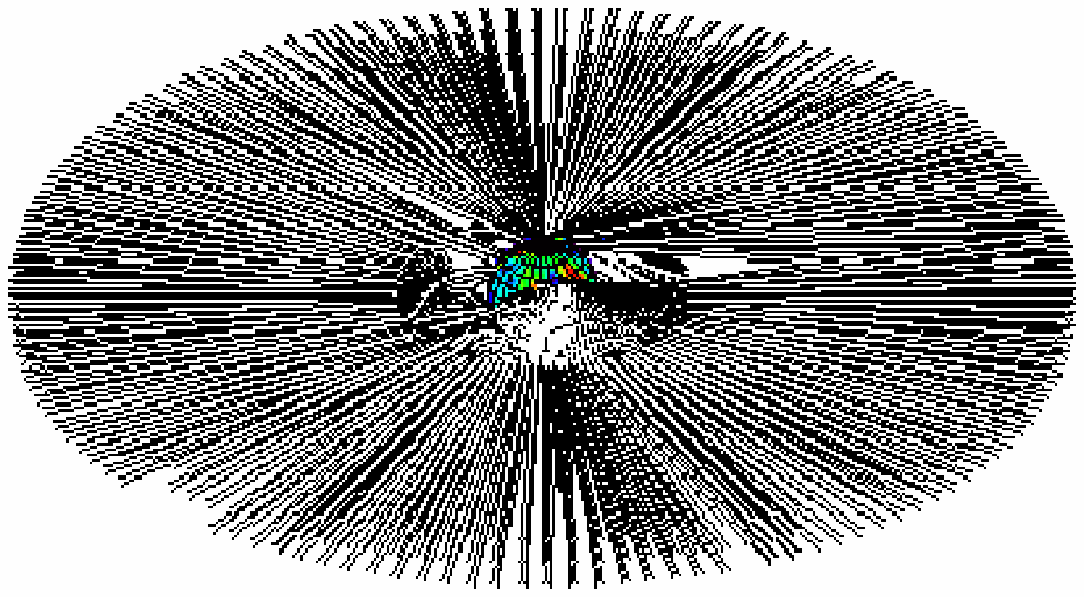,width=80mm}
                        }  \\
        \end{tabular} 
        \caption[]{Field lines which could support accretion flows for a model of a CTTS with
                   a field topology that resembles (a) LQ Hya, obtained
                   using the DF Tau parameters from Table \ref{table}, and (b) AB Dor using the
                   CY Tau parameters.  The stellar surface is coloured to show the strength of the 
                   radial component of the field, with red representing 1kG and black -1kG.  Field 
                   lines have been drawn from the corotation radius.  For the lower mass star, 
                   DF Tau, the natural extent of its corona would be beyond
                   corotation and therefore there is a mixture of open and closed field lines 
                   threading the disc at $R_{co}$.  The higher mass star, CY Tau, has a more compact
                   corona and material flows along open field lines from corotation. 
                   }
    \label{fields}
\end{figure*}

%-------------------------------------------------------------------

\subsection{Coronal extent}
We determine the maximum possible extent of the corona (which is the extent of the 
source surface) by determining the maximum radius at which a magnetic field could
contain the coronal gas.  Since a dipole field falls off with radius most slowly,
we use this to set the source surface.  For a given surface magnetogram we calculate
the dipole field that has the same average field strength.  We then need to calculate
the hydrostatic pressure along each field line.  For an isothermal corona and assuming 
that the plasma along the field is in hydrostatic equilibrium then,
\begin{equation}
  p_s=p_0\exp{\left(\frac{1}{c_s^2}\oint_{s}g_sds \right)}, 
  \label{phydro}
\end{equation}
where $c_s$ is the isothermal sound speed and $g_s$ the component of the effective
gravity along the field line such that, $g_s=\mathbf{g} \cdot \mathbf{B}/|\mathbf{B}|$, $p_0$
is the gas pressure at a field line foot point and $p_s$ the pressure at some point along 
the field line.  The effective gravity in spherical coordinates for a star with 
rotation rate $\omega$ is,
\begin{equation}
  \mathbf{g}\left( r,\theta,\phi \right) = 
         \left ( -\frac{GM_{\ast}}{r^2}+\omega^2r\sin^2{\theta},
         \omega^2r\sin{\theta}\cos{\theta},0 \right). 
  \label{gravity}
\end{equation}
We can then calculate how the plasma $\beta$, the ratio of gas to magnetic pressure, 
changes along each field line.  If at any point along a field line $\beta >1$ then we
assume that the field line is blown open.  This effect is incorporated into our model by 
setting the coronal (gas) pressure to zero whenever it exceeds the magnetic 
pressure ($\beta >1$).  We also set the coronal pressure to zero for open field 
lines, which have one foot point on the star and one at infinity.  The gas pressure, 
and therefore the plasma $\beta$, is dependent upon the choice of 
$p_0$ which is a free parameter of our model.  \citet{jar06} provide a detailed explanation of 
how $p_0$, the coronal base (gas) pressure, can be scaled to the magnetic pressure at a field 
line foot point, so we provide only an outline here. We assume that the base pressure is 
proportional to the magnetic pressure then $p_0 = KB_0^2$, a technique which has been used 
successfully to calculate mean coronal densities and X-ray emission measures for the Sun and 
other main sequence stars (\citealt{jar02a,jar02b}).  By varying the 
constant $K$ we can raise or lower the overall gas pressure along field line loops.  
If the value of $K$ is large many field lines would be blown open and the corona would be 
compact, whilst if the value of $K$ is small then the magnetic field is able to contain more 
of the coronal gas.  The extent of the corona therefore depends both on the value of $K$ 
and also on $B_0$ which is determined directly from surface magnetograms.  For an observed 
surface magnetogram the base magnetic pressure $B_0$ varies across the stellar surface, and as 
such so does the base pressure $p_0$ at field line foot points.  By considering stars from the 
COUP dataset \citet{jar06} obtain the value of $K$ which results in the best fit to 
observed X-ray emission measures, for a given surface magnetogram (see their Table 1).  
We have adopted the same values in this paper.  We then make a conservative estimate of 
the size of a star's corona by calculating the largest radial distance at which a dipole field 
line would remain closed, which we refer to as the {\it source surface} $R_s$.

%----------------------------------------------------------------------------------

\subsection{Coronal stripping}
Lower mass stars, which have small surface gravities and therefore large pressure scale
heights, typically have more extended coronae which would naturally extend beyond
the corotation radius.  Closed field lines threading the disc beyond corotation would 
quickly be wrapped up and sheared open.  Therefore if the maximum extent of a star's
corona is greater than the corotation radius then we set it to be the corotation radius 
instead.  Higher mass stars, with their larger surface gravities, typically have more 
compact coronae which may not extend as far as the corotation radius.  Therefore if
we assume that the disc is truncated at corotation, then accretion proceeds from the inner 
edge of the disc along radial open field lines.  \citet{jar06} provide a discussion
about the extent of T Tauri coronae relative to corotation radii.  It is also worth 
noting that \citet{saf98} criticises current magnetospheric accretion models 
for not including the effects of a stellar corona.  He argues that the inclusion of a 
realistic corona blows open most of the closed field with the eventual net effect being 
that the disc would extend closer to the star.  However, in our model there are open field
lines threading the disc at corotation and it is therefore reasonable to assume that they 
are able to carry accretion flows. 

\begin{table}
  \caption{Data for CTTSs from \citet{val04}}
  \begin{tabular}{ccccc}
  \hline
    $Star$ & $M_{\ast}(M_{\odot})$ & $R_{\ast}(R_{\odot})$ & $P_{rot}(d)$ & $R_{co}(R_{\ast})$\\   
  \hline
    DF Tau & 0.17 & 3.9 & 8.5 & 2.47\\
    CY Tau & 0.58 & 1.4 & 7.5 & 9.55\\
  \hline
\end{tabular}
\label{table}
\end{table}

%------------------------------------------------------------------------------

\subsection{Field extrapolations}
The initial field extrapolation yields regions of open and closed field lines.  Of these,
some intersect the disc and may be actively accreting.  For closed field lines which do 
not intersect the disc it is possible to calculate the X-ray emission measure in the 
same way as \citet{jar06}.  To determine if a field line can accrete we find where it
threads the disc and calculate if the effective gravity along the path of the 
field line points inwards, towards the star.  From this subset of field lines we select those 
which have $\beta <1$ along their length. In other words, for any given solid angle we assume 
that accretion can occur along the first field line within the corotation radius which is able to 
contain the coronal plasma. We assume that the loading of disc material onto the field lines is 
infinitely efficient, such that the first field line at any azimuth which satisfies the accretion 
conditions will accrete, and that field lines interior to this are shielded from the accretion flow.  
We also assume that the accreting field is static and is therefore not distorted by the disc or by the 
process of accretion.  In \S3.3 we consider in more detail how to determine which field lines 
are able to support accretion flows, in order to calculate mass accretion rates and accretion 
filling factors.

Fig. \ref{fields} shows the first set of field lines which may be accreting, obtained by 
surrounding the field extrapolations of LQ Hya and AB Dor with a thin wedge-shaped accretion disc,
with an opening angle of approximately 10\degr.  In \S3
we develop a model for isothermal accretion flows where material leaves the disc
at a low subsonic speed, but arrives at the star with a large supersonic speed.  Not all
of the field lines in Fig. \ref{fields} are capable of supporting such accretion flows, and 
instead represent the maximum possible set of field lines which may be accreting.  We assume a 
coronal temperature of 10MK and obtain the gas pressure at the base of each field line as discussed 
in \S2.1 and by \citet{jar06}.  The natural extent of the corona of DF Tau would be beyond the corotation 
radius and therefore accretion occurs along a mixture of closed and open field lines from 
corotation.  One suggestion for how accretion may proceed along open field lines is that an 
open field line which stretches out into the disc, may reconnect with another open field line 
for long enough for accretion to occur, only to be sheared open once again.  This is of 
particular importance for the higher mass stars, such as CY Tau, where in some cases we find 
that the inner edge of the disc is sitting in a reservoir of radial open field lines.  This may 
have important implications for the transfer of torques between the disc and star.  However 
more work is needed here in order to develop models for accretion along open field lines.

These field extrapolations suggest that accretion may occur along field lines that have very
different geometries.  Indeed, a substantial fraction of the total mass accretion rate may
be carried on open field lines.  Before developing a detailed model of the mass accretion
process, however, we first consider the simple case of a tilted dipole.  While this is an 
idealisation of the true stellar field it allows us to clarify the role that the geometry of
the field may have in governing the mass accretion process.

%------------------------------------------------------------------------------------------

\section{Accretion to a dipole}
\begin{figure}
\centering
\psfig{file=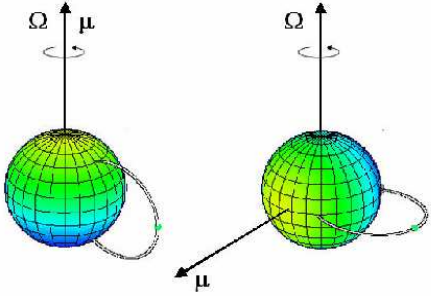,width=8.5cm,angle=0}
\caption{An aligned and tilted dipole field geometry.  The aligned dipole
         (left) with a field line in the star's meridional plane, with the dipole moment 
          $\bmu$ aligned with the stellar rotation axis $\mathbf\Omega$, and the perpendicular dipole 
         (right) with a field line in the star's equatorial plane, with $\bmu$ perpendicular to 
         $\mathbf\Omega$.  The average surface field strength matches that considered by
         \citet{jar06} with yellow (blue) representing the positive (negative) magnetic pole.}
\label{dipoles}
\end{figure}

We have constructed two simple analytic models as sketched in Fig. 
\ref{dipoles}. 
The first case is for a star with a dipolar field with the dipole moment 
$\bmu$ 
aligned with the stellar rotation axis $\mathbf\Omega$.  In standard 
spherical 
coordinates this field may be described as,
\begin{equation}
\mathbf{B}=\left(\frac{2\mu}{r^3}\cos{\theta}, 
\frac{\mu}{r^3}\sin{\theta},0 \right),
\end{equation}
a scenario that allows us to model accretion flows along
field lines in the star's meridional plane.  If we then take this
field structure and tilt it by $\pi/2$ radians such that $\bmu$ now
lies in the star's equatorial plane, perpendicular to $\mathbf\Omega$,
then those field lines which ran north-south in the meridional
plane, now lie east-west in the equatorial plane, with,
\begin{equation}
\mathbf{B}=\left(\frac{2\mu}{r^3}\cos{\phi},0,\frac{\mu}{r^3}\sin{\phi} 
\right).
\label{equfield}
\end{equation}  
Throughout we shall refer to these cases as the \emph{perpendicular dipole} 
for the tilted dipole field and the \emph{aligned dipole} for the aligned 
dipole field.  To do this we consider steady
isothermal accretion flows from a thin accretion disc oriented such that
the disc normal is parallel to the stellar rotation axis.  An
initial sonic Mach number, ${\cal M}$, is ascribed to the accreting material.
We then calculate the pressure and velocity profiles,
relative to arbitrary initial conditions defined at the disc
plane. We calculate the ratio of pressure $p$ at each point
along a field line, relative to that at the disc, $p_d$; and then from this 
we calculate how the Mach number of the flow changes along the field.

The path of a field line may be described by
\begin{equation}
\frac{B_r}{dr}=\frac{B_{\theta}}{rd\theta}=\frac{B_{\phi}}{r\sin{\theta}d\phi}.
\end{equation}
For the perpendicular dipole, where $\theta=\pi/2$ for all field lines that pass through
the disc, it is quickly established
that,
\begin{equation}
\sin^2{\phi} = \Psi r,
\end{equation}
where $\Psi$ is a constant along a particular field line, such that 
different values of $\Psi$ correspond to different field lines.  For the 
perpendicular dipole case, at $r=R_d$, the maximum radial extent of the field line, 
$\phi=\pi/2$; thus, $\Psi=1/R_d$. Using this result the magnitude of the 
magnetic field at a point along a field line a distance $r$ from the centre 
of the star, relative to that at the disc, may be obtained from 
(\ref{equfield}),
\begin{equation}
\frac{B}{B_d} = \left (\frac{R_d}{r}\right )^3 \sqrt{\left ( 
4-\frac{3r}{R_d} \right )}, 
\label{bfield}
\end{equation}
where $B_d=B \left( r=R_d \right )$.  An identical expression can be 
derived for the aligned dipole case.

%---------------------------------------------------------------------------

\subsection{Steady isothermal accretion flows}
The momentum equation for a steady inviscid flow along a flux tube
is
\begin{equation}
\rho \left ( \mathbf{v} \cdot \nabla \right )\mathbf{v}
     =-\nabla \left (p+\frac{B^2}{2\mu} \right ) +\frac{1}{\mu}
      \left ( \mathbf{B} \cdot \nabla \right )\mathbf{B} +
      \rho \mathbf{g}_{eff}-2\rho \mathbf{\bomega} \times \mathbf{v},
\end{equation}
where the symbols have their usual meaning with $\mathbf{g}_{eff}$ being 
the sum of the gravitational and centrifugal accelerations.  In a frame 
of reference rotating with the the star the effective gravity is,
\begin{equation}
\mathbf{g}_{eff} = \mathbf{g} - \mathbf{\bomega} \times \left
(\mathbf{\bomega} \times \mathbf{r} \right ).
\end{equation}
The component of the momentum equation along a field line is then,
\begin{equation}
\rho \frac{d}{ds} \left ( \frac{v^2}{2}\right ) = -\frac{dp}{ds}
      + \rho \mathbf{g}_{eff} \cdot \mathbf{\hat{s}},
\label{momentum_equ}
\end{equation}
since the Coriolis term ($-2\rho \bomega \times
\mathbf{v}$) does not contribute for flows along the field, and where
$\mathbf{\hat{s}}(\mathbf{r})$ is a unit vector along the path of the 
field line.  
Throughout terms with a subscript $d$ will denote quantities defined at 
the disc; 
for example $\rho_d, p_d, v_d$ and $B_d$ are respectively the density,
pressure, velocity along the field and the magnetic field strength as 
defined at 
the plane of the disc, a radial distance $R_d$ from the centre of the 
star.  
Integrating equation (\ref{momentum_equ}) from the disc plane to some 
position 
along the field line at a distance $r$ from the stellar centre, and 
using the 
isothermal equation of state for an ideal gas, $p=\rho c_s^2$, gives,
\begin{equation}
\ln{\left (\frac{p}{p_d}\right )} = \frac{1}{c_s^2}
         \left [ - \frac{1}{2} \left( v^2-v_d^2 \right )
         +\int \mathbf{g}_{eff} \cdot \mathbf{\hat{s}}ds\right ],
\label{pressure}
\end{equation}
where $c_s$ is the isothermal sound speed.  If we assume that both mass 
and magnetic 
flux are conserved along each flux tube (of cross-sectional area $A$), 
then the flow 
must satisfy
\begin{eqnarray}
\frac{d}{ds} \left ( \rho v A \right ) &=& 0 \\
\frac{d}{ds} \left ( B A \right ) &=& 0,
\end{eqnarray}
which may be expressed equivalently as,
\begin{equation}
\frac{pv}{B} = \frac{p_d v_d}{B_d} = const, 
\label{constant}
\end{equation}
By combining (\ref{pressure}) with (\ref{constant}) a relation for the 
pressure 
structure along an accreting field line can be established,
\begin{equation}
\ln{\left (\frac{p}{p_d}\right )} + \frac{1}{2}{\cal M}^2 \left (
\frac{p_d}{p}\frac{B}{B_d}\right )^2
          -\frac{1}{2}{\cal M}^2 - \frac{1}{c_s^2}\int 
\mathbf{g}_{eff} \cdot \mathbf{\hat{s}}ds = 0,
\label{pre}
\end{equation}
and also directly from (\ref{constant}) an expression for the
velocity structure,
\begin{equation}
\frac{v}{c_s} = {\cal M} \frac{p_d}{p} \frac{B}{B_d}, \label{velo}
\end{equation}
where in both cases ${\cal M}=v_d/c_s$ denotes the initial sonic Mach number 
at which the 
accretion flow leaves the plane of the disc. It is then possible to find 
the pressure 
at each point along a field line, relative to the pressure at the disc 
($p/p_d$), by
finding the roots of (\ref{pre}).  Once these roots have been found the 
velocity profile 
can be obtained from (\ref{velo}) by calculating how the Mach number of 
the flow varies as material moves from the disc to the star.

%------------------------------------------------------------------

\subsection{Pressure and velocity profiles}

\begin{figure*}
        \def\subfigtopskip{4pt}
        \def\subfigbottomskip{4pt}
        \def\subfigcapskip{2pt}
        \centering
        \begin{tabular}{cc}
        \subfigure[]{
                        \label{prefig}                  
                        \psfig{figure=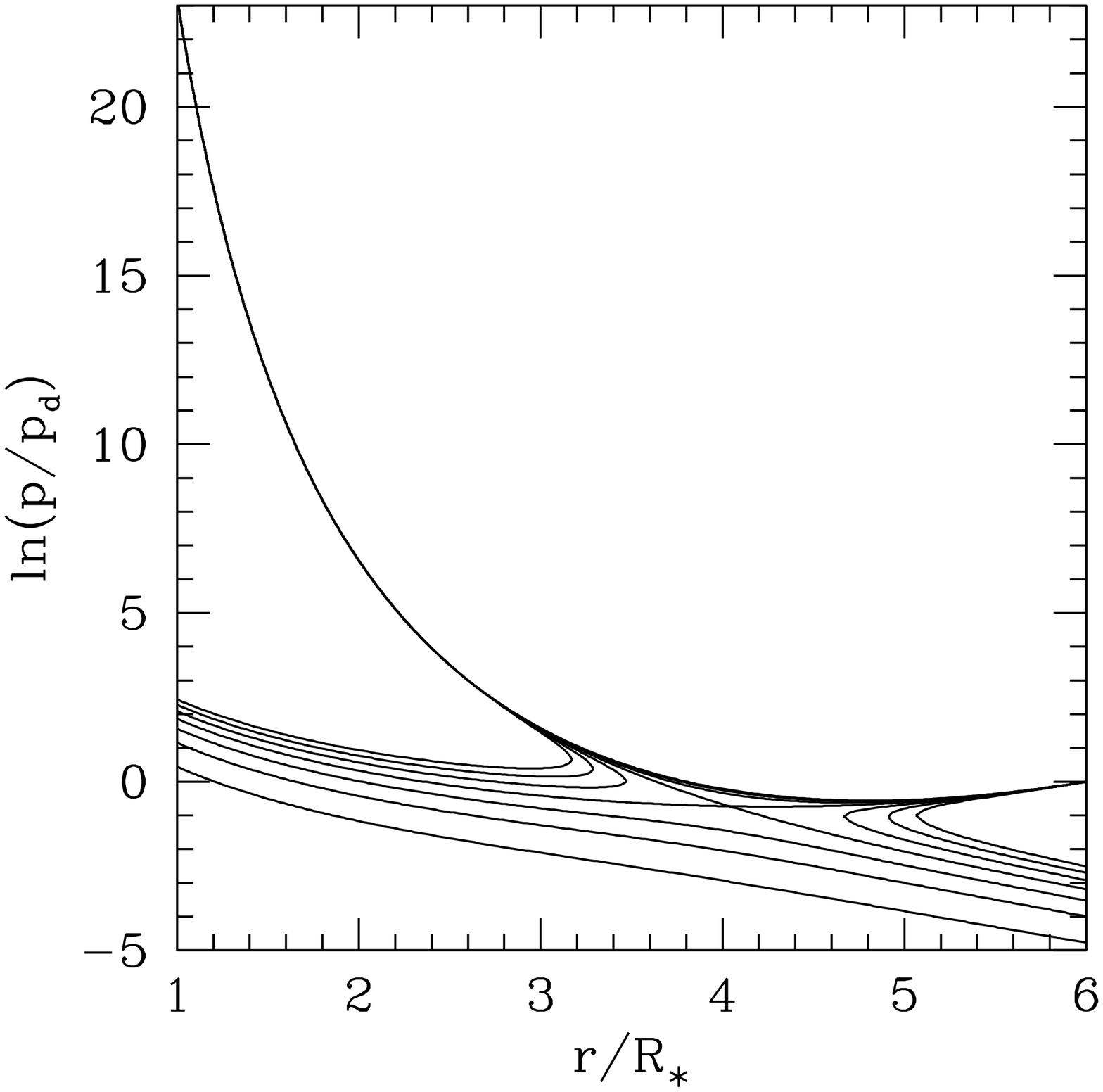,width=80mm}
                        } &
                \subfigure[]{
                        \label{velofig} 
                        \psfig{figure=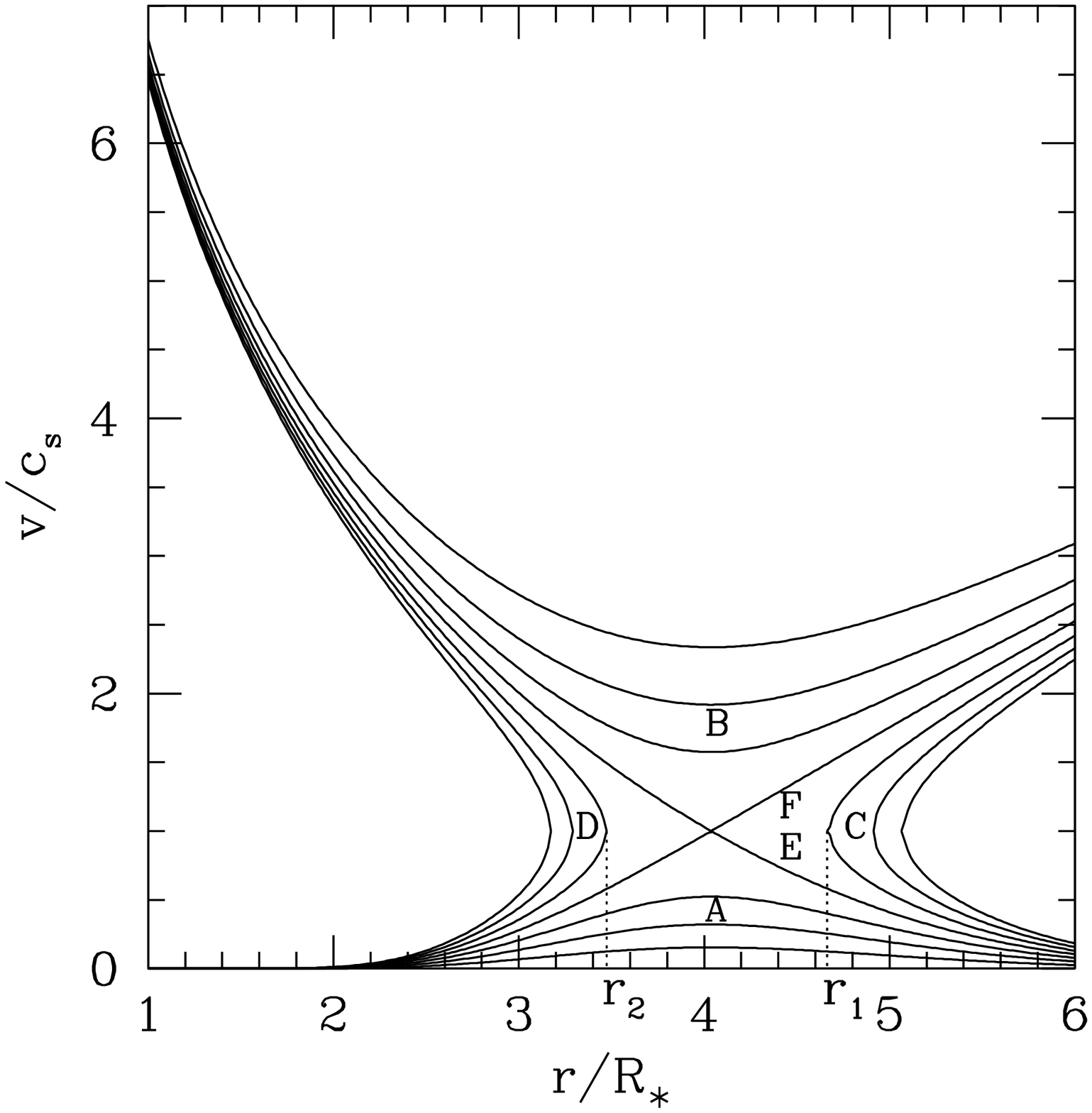,width=80mm}
                        }  \\
        \end{tabular} 
        \caption[]{The resulting pressure and velocity profiles for accretion along
                   equatorial dipole field lines.  The inner edge of the 
                   disc is at $R_{d} = 6.0R_{\ast}$ which is approximately the 
                   corotation radius.  Different lines represent different initial velocities.}
    \label{profiles}
\end{figure*}

For the perpendicular dipole, in the star's equatorial plane, the effective gravity 
has only a radial component,
\begin{equation}
\mathbf{g}_{eff} = \left ( -\frac{GM_{\ast}}{r^2}+\omega^2 r \right )
\mathbf{\hat{r}}.
\label{grav_equ}
\end{equation}
Taking the component of the effective gravity along the field, that is 
along a path 
parameterised by $\mathbf{\hat{s}}=\mathbf{B}/B$, substituting into 
(\ref{pre}),
and using (\ref{bfield}) gives an expression for the pressure structure 
along equatorial
field lines,
\begin{eqnarray}
\ln {\left ( \frac{p}{p_d}\right )}+\frac{1}{2} {\cal M}^2 
\left(\frac{R_d}{r}\right )^6 \left ( 
                4-\frac{3r}{R_d}\right ) \left ( \frac{p_d}{p} \right 
)^2-\frac{1}{2} {\cal M}^2 
                \nonumber \\
                + \Phi_g \left (\frac{1}{R_d}-\frac{1}{r} \right )
                +\Phi_c \left ( R_d^2 - r^2 \right )= 0,
\label{equatorial}
\end{eqnarray}
where both $r$ and $R_d$ are measured in units of the stellar radius
$R_{\ast}$ and $\Phi_g$ and $\Phi_c$ are the surface ratios 
of the gravitational and centrifugal energies to the thermal energies,
\begin{eqnarray}
\Phi_g &=& \frac{GM_{\ast}}{R_{\ast}c_s^2} \\
\Phi_c &=& \frac{1}{2} \left( \frac{\omega R_{\ast}}{c_s} \right )^2.
\end{eqnarray}
The roots of (\ref{equatorial}) give the pressure at some point along a 
field line loop which is a radial distance $r$ from the stellar centre.

For the aligned dipole, in the star's meridional plane, the effective gravity 
has both an $r$ and $\theta$ component,
\begin{equation}
\mathbf{g}_{eff} = \left ( -\frac{GM_{\ast}}{r^2} + \omega^2 r
\sin^2{\theta} \right )\mathbf{\hat{r}}
                   + \left( \omega^2 r \sin{\theta}\cos{\theta} 
\right) \hat{\btheta}.
\label{grav_perp}
\end{equation}
Following an identical argument to that above it can be
established that for accretion in the star's meridional plane, the 
pressure function 
(\ref{pre}) becomes,
\begin{eqnarray}
\ln{\left ( \frac{p}{p_d} \right )}+\frac{1}{2} {\cal M}^2 
\left(\frac{R_d}{r}\right )^6 
           \left ( 4-\frac{3r}{R_d}\right )\left ( \frac{p_d}{p} \right 
)^2-\frac{1}{2} {\cal M}^2
           \nonumber \\ 
           + \Phi_g  \left (\frac{1}{R_d}-\frac{1}{r} \right )
           +\Phi_c \left ( R_d^2 - \frac{r^3}{R_d} \right )= 0.
\label{meridional}
\end{eqnarray}
The only difference from the perpendicular dipole is in the final term.

For a CTTS with a mass of 0.5M$_{\odot}$, radius 2R$_{\odot}$ and a
rotation period of 7 days we have calculated the pressure and
velocity structure along accreting dipole field lines, for a range
of accretion flow temperatures, starting radii and initial sonic Mach numbers.
Figs. \ref{prefig} and \ref{velofig} show a typical pressure and velocity profile
for the perpendicular dipole, whilst those for the aligned dipole are qualitatively similar.
The pressure profile shows how the ratio $p/p_d$, where $p$ is the pressure along the 
field line and $p_d$ the pressure at the disc, varies as the flow moves from the disc to the 
star (plotted logarithmically for clarity).  The velocity profile shows how the Mach number 
of the flow changes along the field line.  For different accretion flow temperatures and 
starting radii the resulting profiles are similar, except in a few select cases, as discussed 
in the next section.  Fig. \ref{profiles} is for an accretion flow 
leaving the disc at $R_d=6.0R_{\ast}$, which is approximately the equatorial corotation 
radius $R_{co}$ where,
\begin{equation}
R_{co} = \left( \frac{GM_{\ast}}{\omega^2}\right )^{1/3},
\label{corot}
\end{equation} 
and for an accretion flow temperature of 10$^5$K.  This is at least one order of magnitude higher
than what is believed to be typical for accretion in CTTSs, but a higher temperature 
has been selected here in order to clearly illustrate the various types of solutions
labelled in Fig. \ref{velofig}.  At lower temperatures the form of the velocity
solutions is similar.

At the critical radius $r_c$ either the flow velocity equals the sound speed,
$v=c_s$, or $dv/ds=0$.  There are several distinct velocity
solutions labelled in Fig. \ref{velofig}.  For very small subsonic initial
Mach numbers (curve A) the flow remains subsonic all the way to
the star, and for large supersonic initial Mach numbers, it remains
supersonic from the disc to the star (curve B).  There is also a
range of initial Mach numbers (curves C) where the flow will
not reach the star, and one value of ${\cal M}$ that results in a transonic 
solution -- where the flow leaves the disc at a subsonic speed and accelerates 
hitting the star at a supersonic speed (curve E).  Observations of
the widths of line profiles suggest that the accreting material
reaches the stellar surface at several hundred kms$^{-1}$, certainly at
supersonic speeds (e.g. \citealt{edw94}).  The velocity profiles indicate that
it is possible to have accretion flows which leave the disc at a low
subsonic velocity but which arrive at the star with a large supersonic velocity.
In Fig. \ref{velofig} the transonic solution arrives at the star with a
Mach number of 6.54, which at a temperature of 10$^5$K corresponds to an 
in-fall velocity of 243kms$^{-1}$.  For a realistic accretion temperature of 
10$^4$K the in-fall velocity is 259kms$^{-1}$.  

Models of funnel flows have been studied 
using an isothermal equation of state \citep{li99} 
and for a polytropic flow \citep{kol02}, with the latter 
demonstrating that transonic accretion flows are only 
possible for a range of starting radii around the corotation 
radius.  MHD simulations by \cite{rom02} also indicate 
that accretion flows can arrive at the star with large supersonic 
velocities.  However, our aim here is to determine whether or not
the magnetic field geometry has any affect on accretion; in 
particular how the field structure affects the mass accretion rate, 
rather than to discuss the types of velocity solutions that we would 
expect to observe.  In Appendix A we discuss an efficient algorithm 
which allows us to determine both the location of the sonic point
on a field line and the initial Mach number required to give a smooth
transonic solution.  This algorithm may be applied to accretion flows
along field lines of any size, shape and inclination, even in the absence
of analytic descriptions of the magnetic field and effective gravity.   

%---------------------------------------------------------------------------------------
\subsection{Mass accretion rates and filling factors}

\begin{figure*}
        \def\subfigtopskip{4pt}
        \def\subfigbottomskip{4pt}
        \def\subfigcapskip{2pt}
        \centering
        \begin{tabular}{cc}
        \subfigure[]{
                        \label{mdotbeta_dipole}                         
                        \psfig{figure=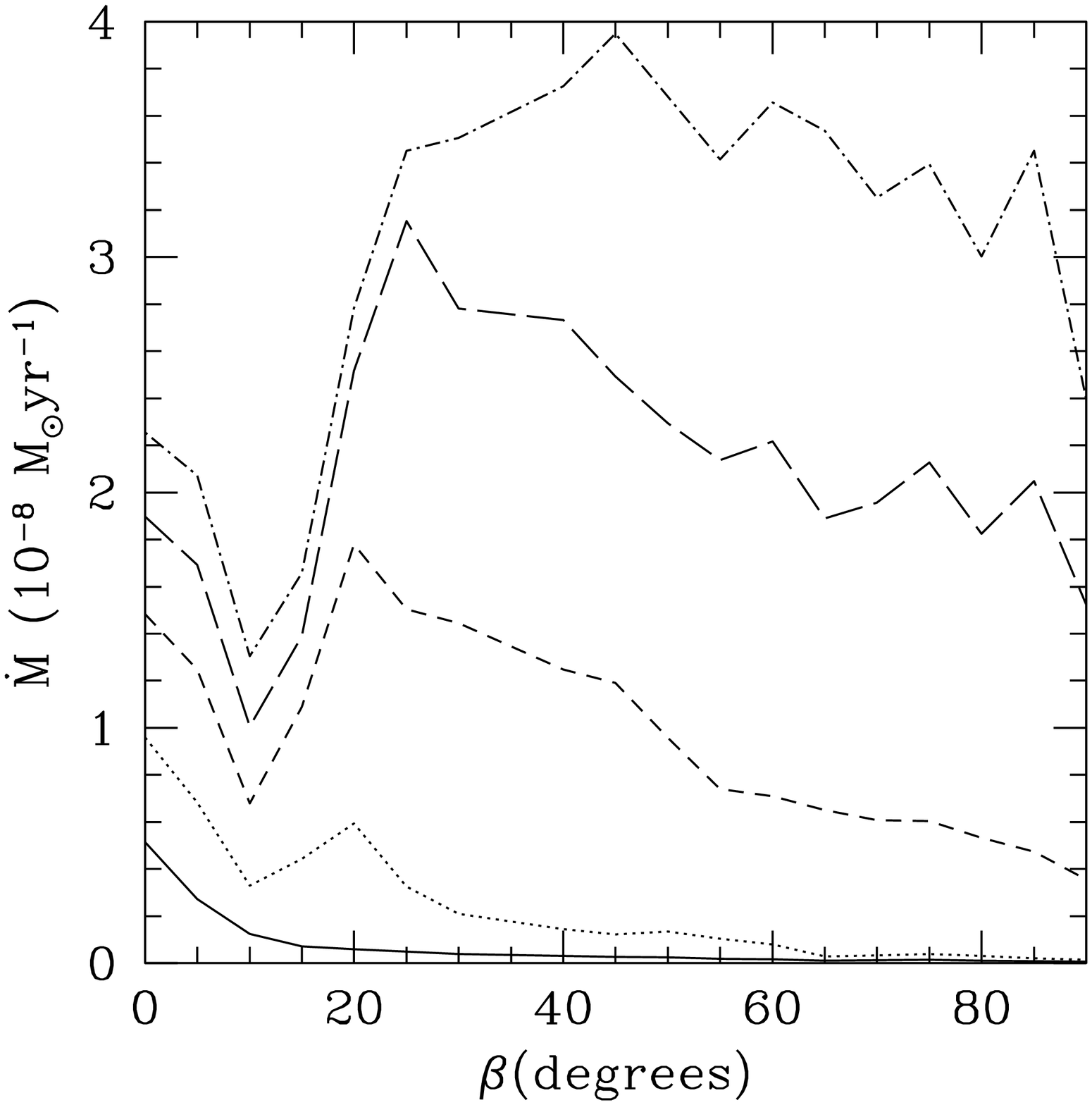,width=80mm}
                        } &
                \subfigure[]{
                        \label{faccbeta_dipole} 
                        \psfig{figure=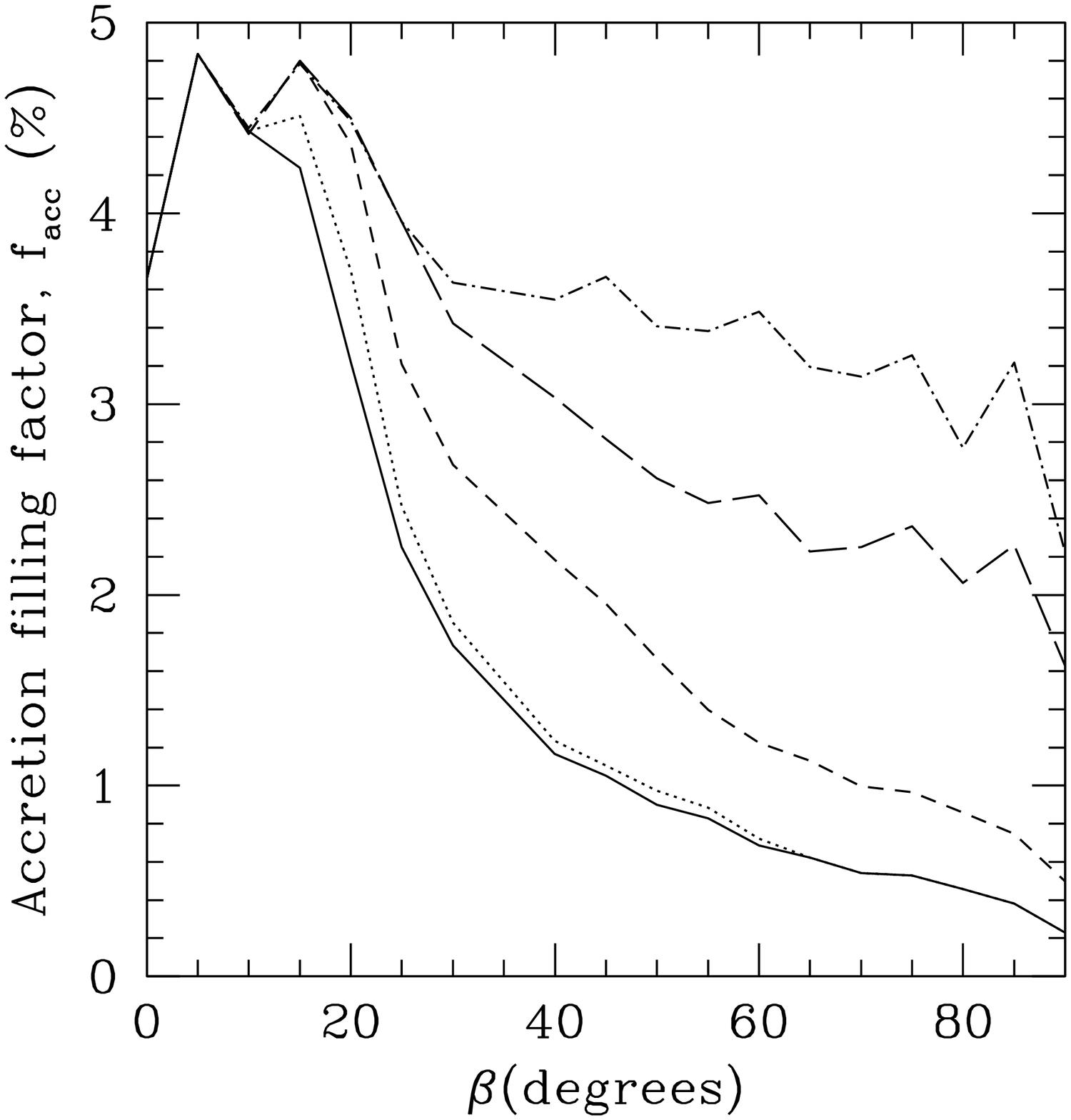,width=80mm}
                        }  \\
                \subfigure[]{
                        \label{closedbeta_dipole} 
                        \psfig{figure=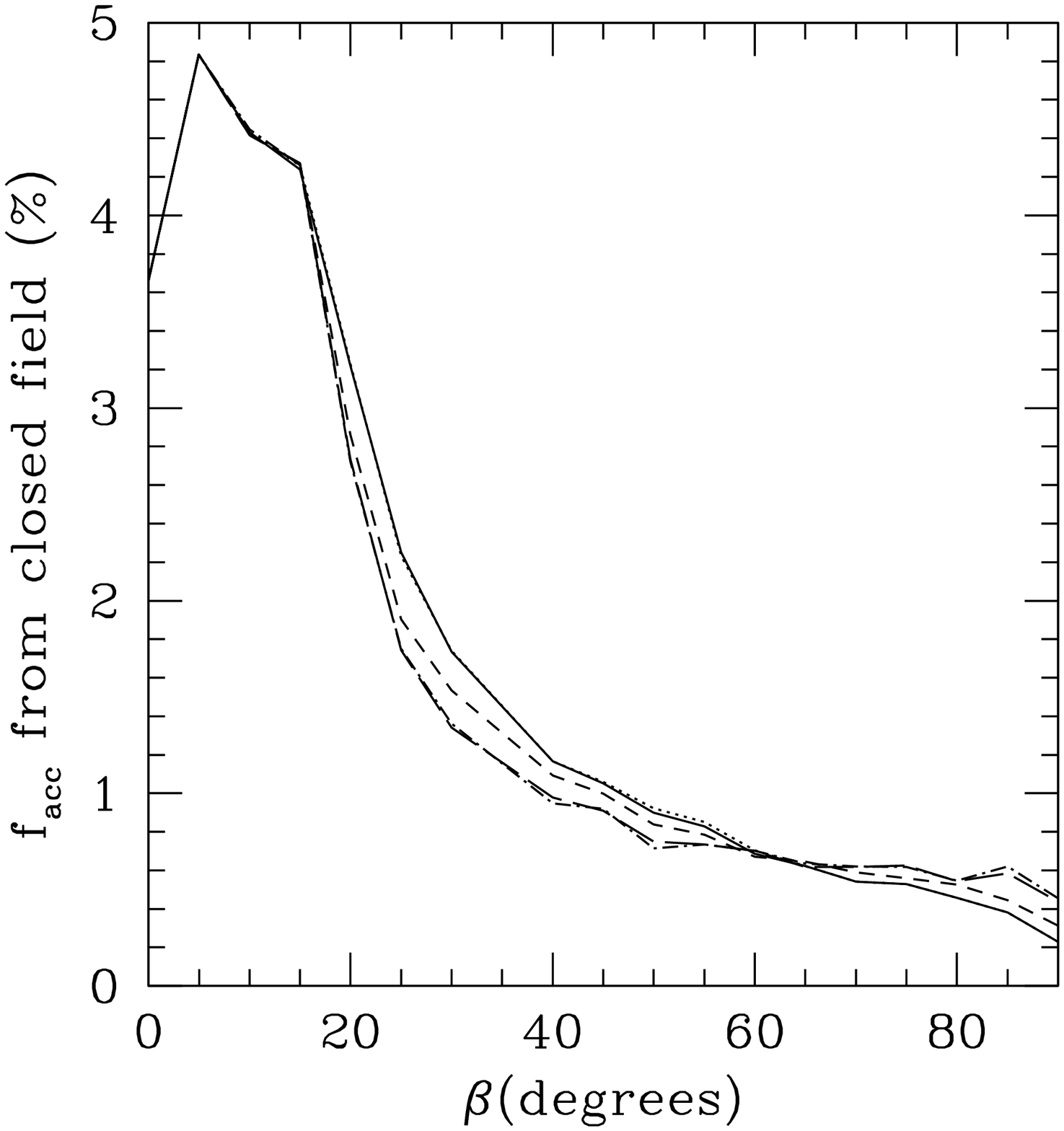,width=80mm}
                        }  &
                \subfigure[]{
                        \label{openbeta_dipole} 
                        \psfig{figure=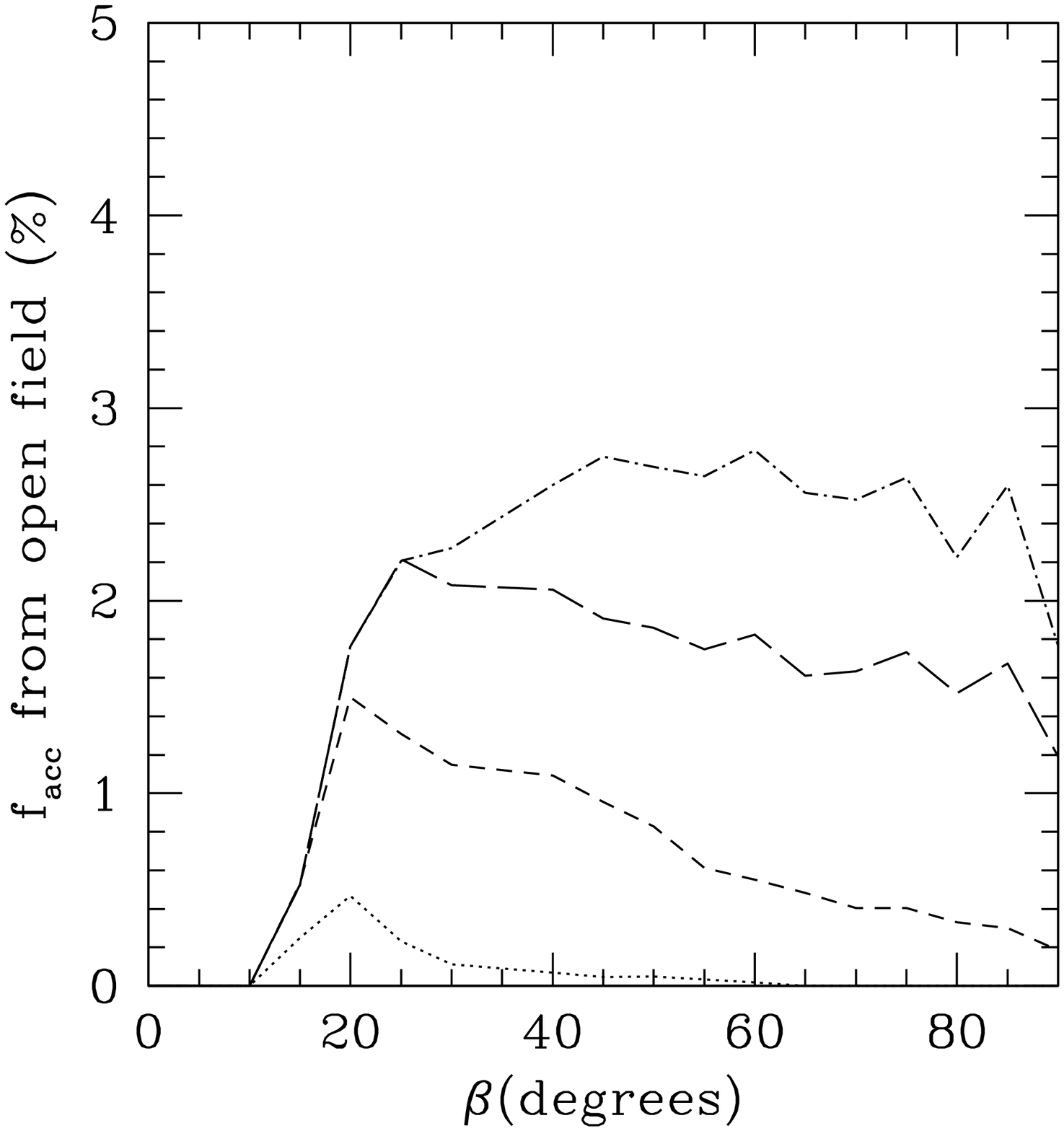,width=80mm}
                        }  \\
        \end{tabular} 
        \caption[] {(a) The change in mass accretion rate and (b) the accretion filling factor as a 
                    function of $\beta$, for accretion to dipole fields where $\beta$ is the obliquity 
                   - the angle between the rotation and magnetic poles, for accretion flow temperatures 
                   of 1000K ({\it solid}), 2500K ({\it dotted}), 5000K ({\it short dash}), 
                   7500K ({\it long dash}) and 10000K ({\it dash-dot}).  (c) and (d) show how the contribution
                   to the accretion filling factor from accreting closed and open field changes with 
                   $\beta$.  The DF Tau parameters from Table \ref{table} have been used.  
                   There are no open accreting field lines for the $T_{acc}=1000K$ case.}
    \label{number}
\end{figure*}

\begin{figure*}
        \def\subfigtopskip{4pt}
        \def\subfigbottomskip{4pt}
        \def\subfigcapskip{2pt}
        \centering
        \begin{tabular}{cc}
        \subfigure[]{
                        \label{dipole_feet0}                         
                        \psfig{figure=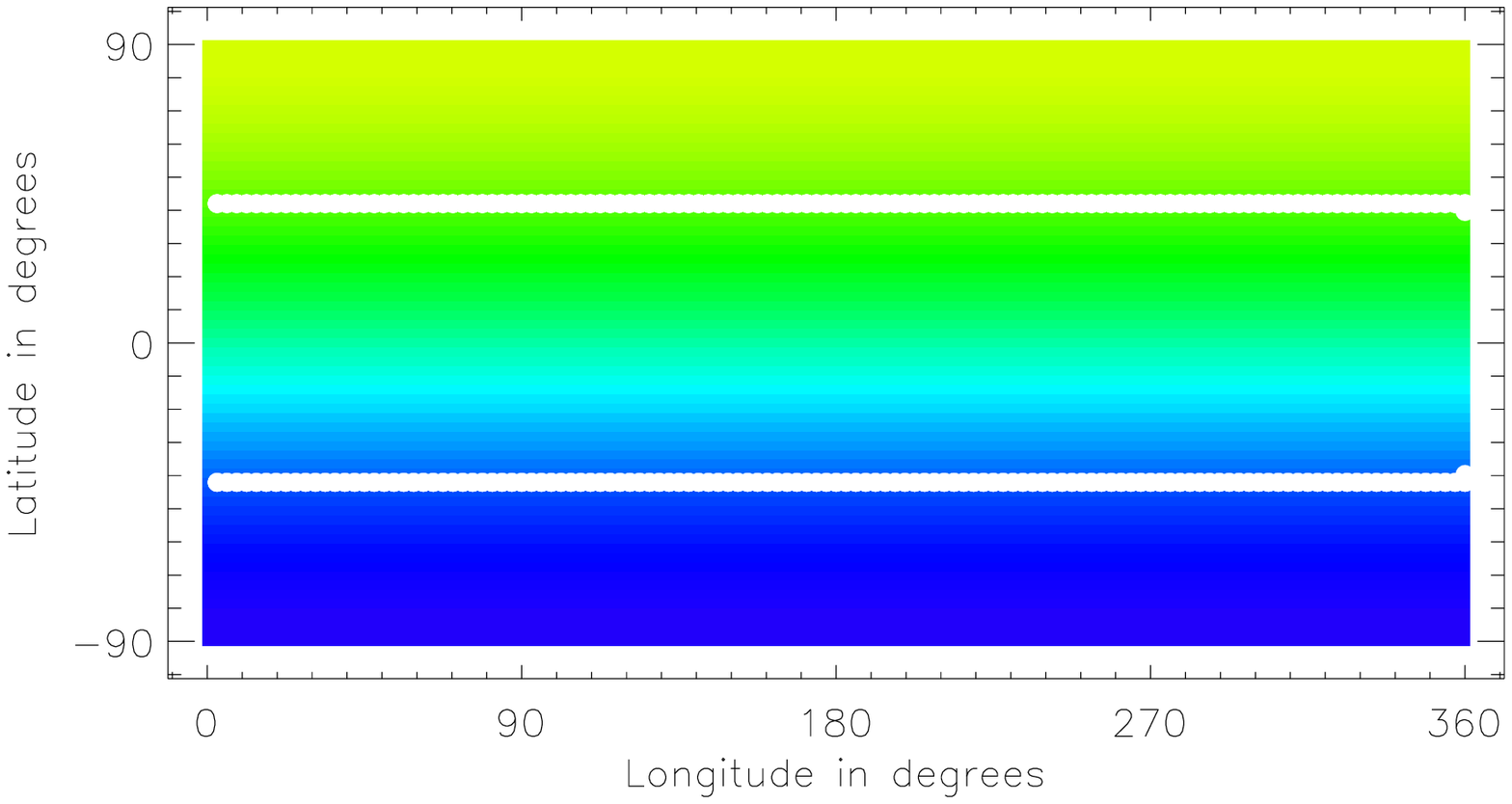,width=80mm}
                        } &
                \subfigure[]{
                        \label{dipole_feet15} 
                        \psfig{figure=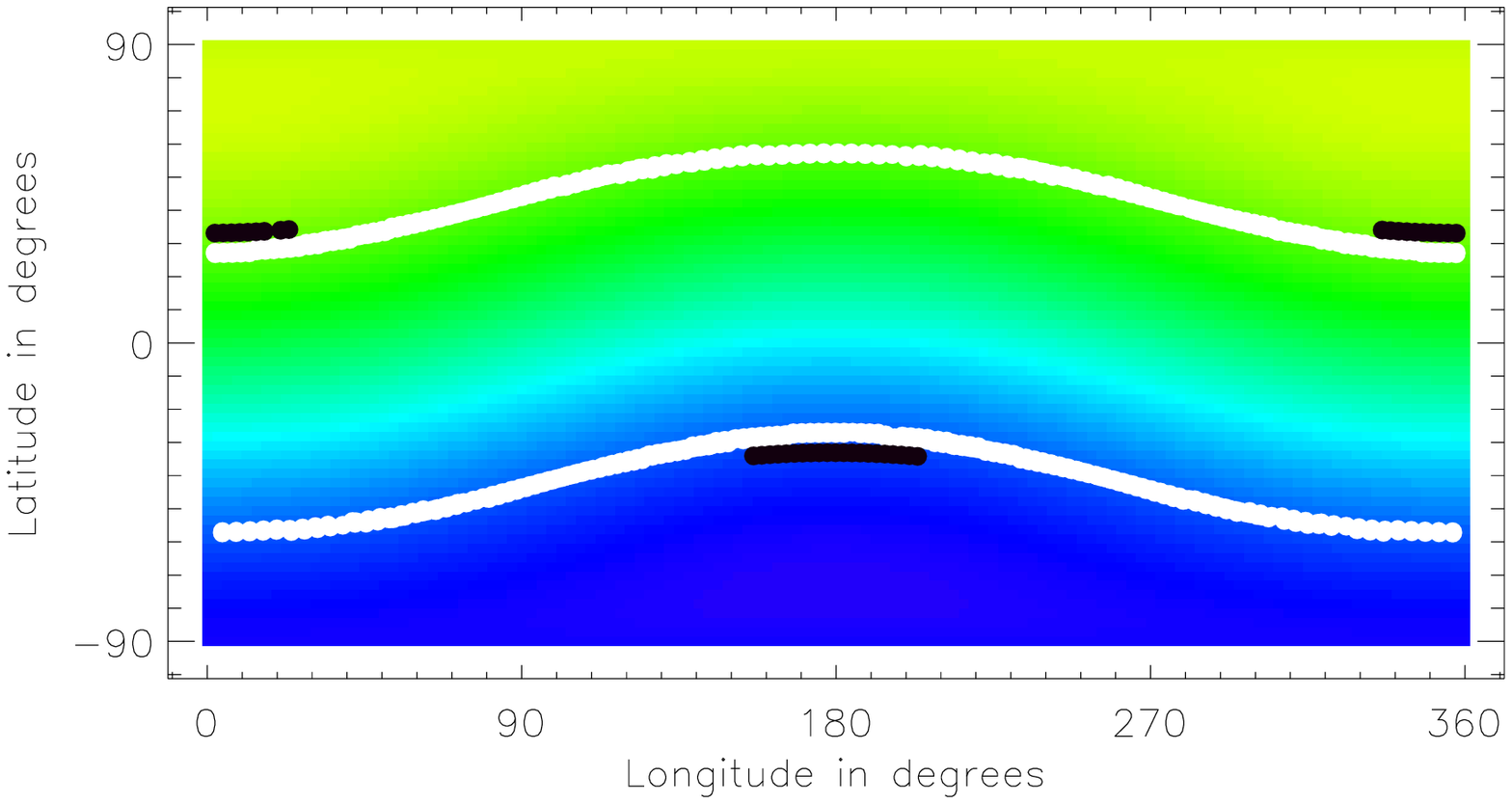,width=80mm}
                        }  \\
                \subfigure[]{
                        \label{dipole_feet45} 
                        \psfig{figure=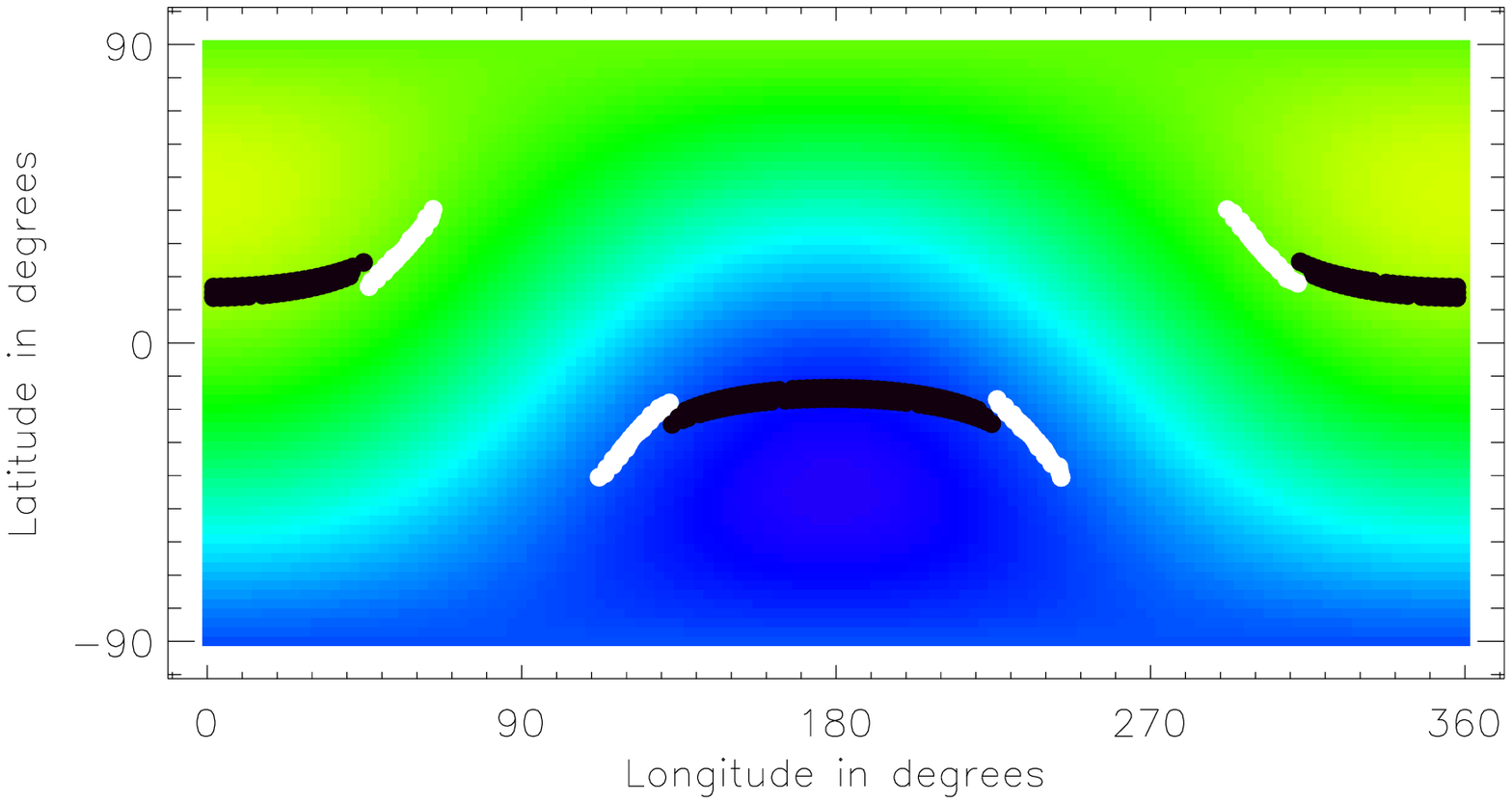,width=80mm}
                        }  &
                \subfigure[]{
                        \label{dipole_feet90} 
                        \psfig{figure=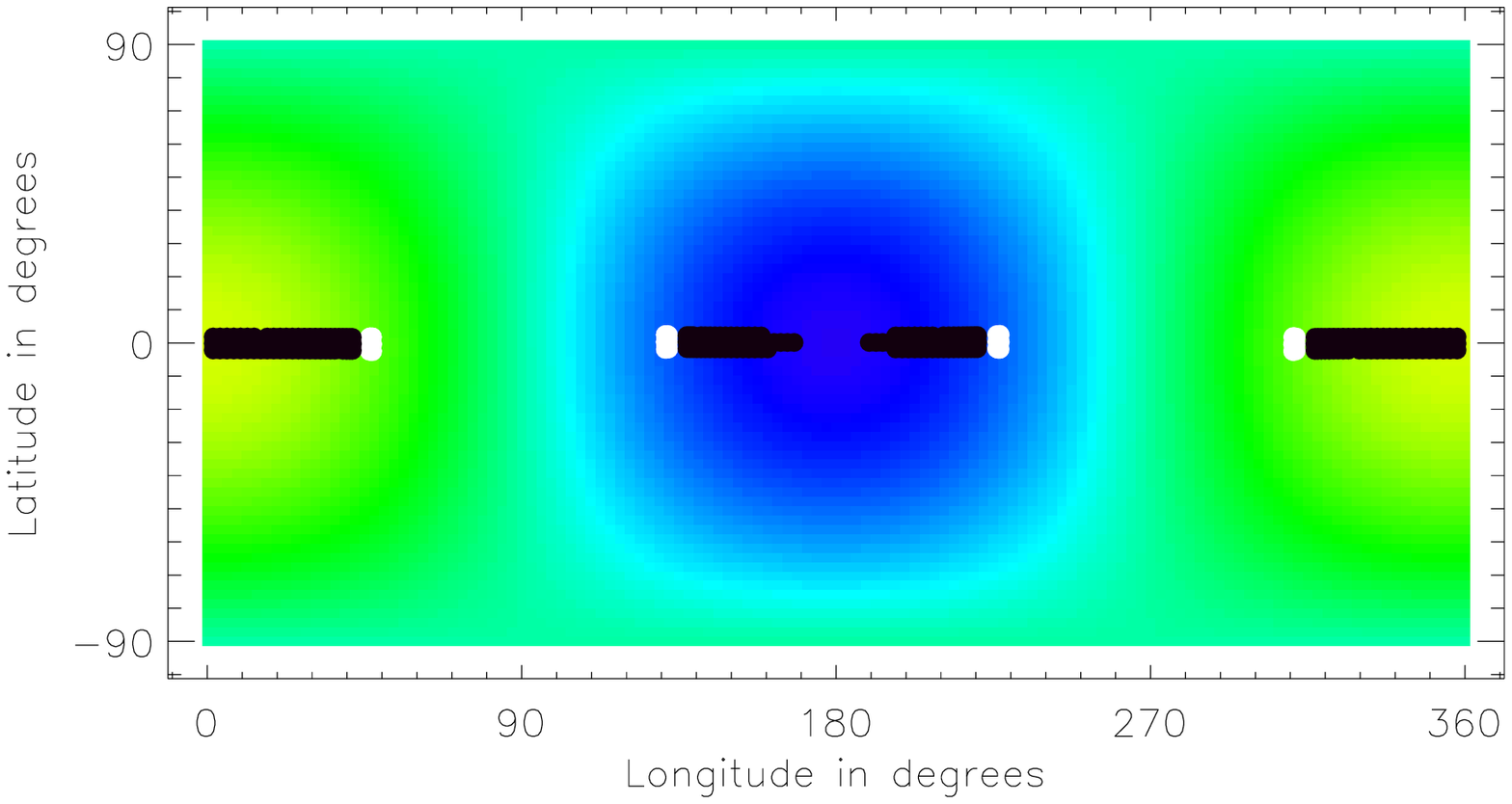,width=80mm}
                        }  \\
        \end{tabular} 
        \caption[]{The stellar surface with white (black) points indicating the closed (open) accreting 
                  field line foot 
                  points for accretion to a dipole with obliquity (a) $\beta=0\degr$, the
                  aligned dipole, where accretion proceeds onto two rings in opposite hemispheres; (b) 
                  $\beta=15\degr$ where the accretion rings have been distorted and open field lines 
                  produce the small bands centred on 180\degr and 360\degr longitude; (c) $\beta=45\degr$ 
                  where accretion occurs predominantly along the open field lines and (d) $\beta=90\degr$, 
                  the perpendicular dipole, where accretion occurs in bars around the star's equator.  
                  All are for an accretion flow temperature of 10$^4$K.  The average surface field 
                  strength matches that considered by \citet{jar06} with yellow (blue) representing the 
                  positive (negative) magnetic pole.}
    \label{dipole_feet}
\end{figure*}

For accretion to occur the effective gravity 
at the point where a field line threads the disc must point inwards towards the star.  
From this subset of field lines we select those which are able to contain the corona 
and support transonic accretion flows.  We further check to ensure that the plasma beta resulting from
accretion remains $<1$ along their length.  Therefore, at any particular azimuth, accretion 
occurs along the first field line at, or slightly within, the corotation radius; the field line
must be able to contain the coronal plasma and have a sonic point along its length.    
In order to determine if a field line can support a transonic accretion flow, the 
first step is to find the pressure and velocity structure its length,
which will be similar to those described in \S3.2.  To do
this we need to determine the initial Mach number that 
would produce an accretion flow.  We can achieve this by determining if a field line
has a sonic point as discussed in Appendix A.

To calculate a mass accretion rate we require the velocity
and density of each accretion flow at the stellar surface, and also the 
surface area of the star covered in hot spots.  For an assumed accretion
flow temperature we determine the initial Mach number required to generate a transonic
velocity profile, along each field line, and determine the in-fall velocity from (\ref{velo}).
At every point along a field line we know the ratio of pressure at that point, to that 
at the disc, $p/p_d$.  For an isothermal equation of state
$p \propto \rho$, so we also know the ratio of densities $\rho/\rho_d$ at 
every point along the accreting field line.  Thus for a given disc midplane density
$\rho_d$, we can estimate the density at the stellar surface $\rho_{\ast}$.  
Throughout we assume a constant disc midplane density of $\rho_d = 5.0\times 10^{-9}gcm^{-3}$,
a resonable value at the corotation radius for T Tauri stars (e.g. \citealt{bos96}).
The mass accretion rate may be expressed in terms of quantities defined
at the disc plane, with $\dot{M} \propto \rho_d$.  Therefore raising or lowering $\rho_d$
directly increases or decreases $\dot{M}$.  We estimate the total surface area of the star 
covered in accretion hot spots by summing the area of individual 
grid cells which contain accreting field line foot points.  
For each grid cell $i$ (of area $A_i$) on the stellar surface we obtain the average in-fall velocity 
$\bar{v}_{\ast}$ and average density $\bar{\rho}_{\ast}$ of material accreting into that cell.  
Most grid cells do not contain any accreting field line foot points and therefore do not 
contribute to the mass accretion rate.  The mass accretion rate is then the sum over all cells $i$ 
containing accreting field line foot points,
\begin{equation}
\dot{M}=\sum_i\dot{M}_i=\sum_i\left [A\bar{v}_{\ast}\bar{\rho}_{\ast}\right ]_i.
\label{massrateequ}
\end{equation}
The mass accretion rate can be expressed equivalently as $\dot{M}=\rho_d v_d A_d$, where $A_d$
is the surface area of the disc that contributes to accretion (which depends 
on the radial extent of accreting field lines within the disc).  Using the surface area of grid 
cells within the disc which contain accreting field lines to estimate $A_d$, we obtain $\dot{M}$ values 
that are comparable to those calculated from (\ref{massrateequ}).  Therefore it makes
little difference which formulation for $\dot{M}$ is used.  The accretion filling factor $f_{acc}$, 
the fractional surface area of the star covered in hot spots, is then calculated from,
\begin{equation}
f_{acc} = \frac{\sum_{i} A_{i}}{4\pi R_{\ast}^2}
\end{equation}  

We have calculated the total mass accretion rate that dipole fields can support for 
isothermal accretion flow temperatures of between 10$^3$K and 10$^4$K; for values of 
$\beta$ from 0\degr to 90\degr, where $\beta$ is the obliquity of the dipole
(the angle between the dipole moment and the stellar rotation axis); for the DF Tau 
parameters in Table \ref{table} and for a coronal temperature of 10MK 
(see Fig. \ref{number}).  $\beta=0\degr$ and $\beta=90\degr$ 
correspond to the aligned and perpendicular dipoles respectively.  

For dipolar accretion we obtain typical mass accretion rates
of $\dot{M} \approx 10^{-9}-10^{-8}M_{\odot}yr^{-1}$.  The mass accretion rate increases 
with temperature in all cases, but  at low accretion 
temperatures $T_{acc}$, there is little difference in $\dot{M}$ for all values of $\beta$ 
(see Fig. \ref{mdotbeta_dipole}).  For higher $T_{acc}$ values, the aligned dipole field can 
support mass accretion rates which are a factor of two times less than those fields with large 
values of $\beta$.  This, in part, can be attributed to the increase in the amount of open 
field lines which thread the disc, and are able to support accretion as $\beta$ is 
increased (see Fig. \ref{openbeta_dipole}).  As the dipole is tilted from $\beta=0\degr$ to 
$10\degr$ the mass accretion rate is reduced (see Fig. \ref{mdotbeta_dipole}).  This can be 
understood by the changing shape of the closed field lines as $\beta$ is increased.  For accretion 
along aligned dipole field lines, accreting material may flow along two identical paths from the 
disc to the star; that is it may accrete either onto the northern, or the southern hemisphere.  
Once the dipole has been tilted through a small angle, the path along the field onto each hemisphere 
changes, with one segment of the closed field line loop being shallower than before and curved towards the
star, and the other being longer.  This longer segment bulges out slightly, so that
material flowing along such field lines follows a path which initially curves away from the 
star, before looping back around to the stellar surface.  This creates a difference in initial
Mach numbers necessary to create transonic accretion flows along the different field line segments,
with the net result that some closed field line segments are no longer able to accrete 
transonically when $\beta=10\degr$ (see Fig. \ref{closedbeta_dipole}).  As the dipole is tilted 
further from $\beta=10\degr$ to $\approx 30\degr-40\degr$ the mass accretion rate increases in all but 
the lowest $T_{acc}$ cases.  This is because once the dipole has been tilted far enough the open field 
lines (those that have foot points at latitudes closer
to the magnetic axis) begin to intersect the disc (see Fig. \ref{openbeta_dipole}).  There are therefore 
more possible paths that material can take from the disc to the star, causing an increase in $\dot{M}$
(again in all but the lower $T_{acc}$ cases).  As $\beta$ is further increased the amount of accreting closed 
field lines continues to reduce, whilst the amount of open field lines threading the disc reaches a maximum, 
and we therefore see a trend of falling mass accretion rates towards the largest values of $\beta$.   

The accretion flow temperature is important in determining whether open field lines are able to
support transonic accretion flows.  From Figs. \ref{closedbeta_dipole} and \ref{openbeta_dipole} it is 
clear that the contribution to accretion from the closed field is constant for all values of $T_{acc}$, 
whereas the contribution from the open field depends strongly on 
$T_{acc}$, with more open field lines accreting at higher accretion flow temperatures.  At the lowest 
accretion flow temperature which we consider (1000K), there are no open field lines able to
support transonic accretion, even for the large values of $\beta$ where there are many such field
lines passing through the disc.  This can be understood as follows.  For transonic accretion a sonic point 
must exist on a field line.  At a sonic point $v=c_s$; applying this to
(\ref{velo}), substituting into (\ref{pre}) and rearranging gives,
\begin{equation}
\frac{1}{v}\frac{dv}{ds} \left ( v^2 - c_s^2 \right ) =
        \mathbf{g}_{eff} \cdot \mathbf{\hat{s}} - \frac{c_s^2}{B}\frac{dB}{ds},
\label{diff}
\end{equation}
from which it can be seen that there exists some critical radius
$r_c$ where either $v=c_s$ or $dv/ds=0$.  Clearly at this critical radius the
two terms on the RHS of (\ref{diff}) must be equal,
\begin{equation}
\frac{c_s^2}{B}\frac{dB}{ds} = \mathbf{g}_{eff} \cdot \mathbf{\hat{s}}, 
\label{sonic}
\end{equation}
where all the terms are evaluated at $r_c$.  It should be noted that (\ref{sonic}) may 
also be obtained by finding the maximum turning point of (\ref{crit}), consistent with 
our argument in Appendix A.

\begin{figure*}
        \def\subfigtopskip{4pt}
        \def\subfigbottomskip{4pt}
        \def\subfigcapskip{2pt}
        \centering
        \begin{tabular}{cc}
        \subfigure[]{
                        \label{lqhya_feet}                      
                        \psfig{figure=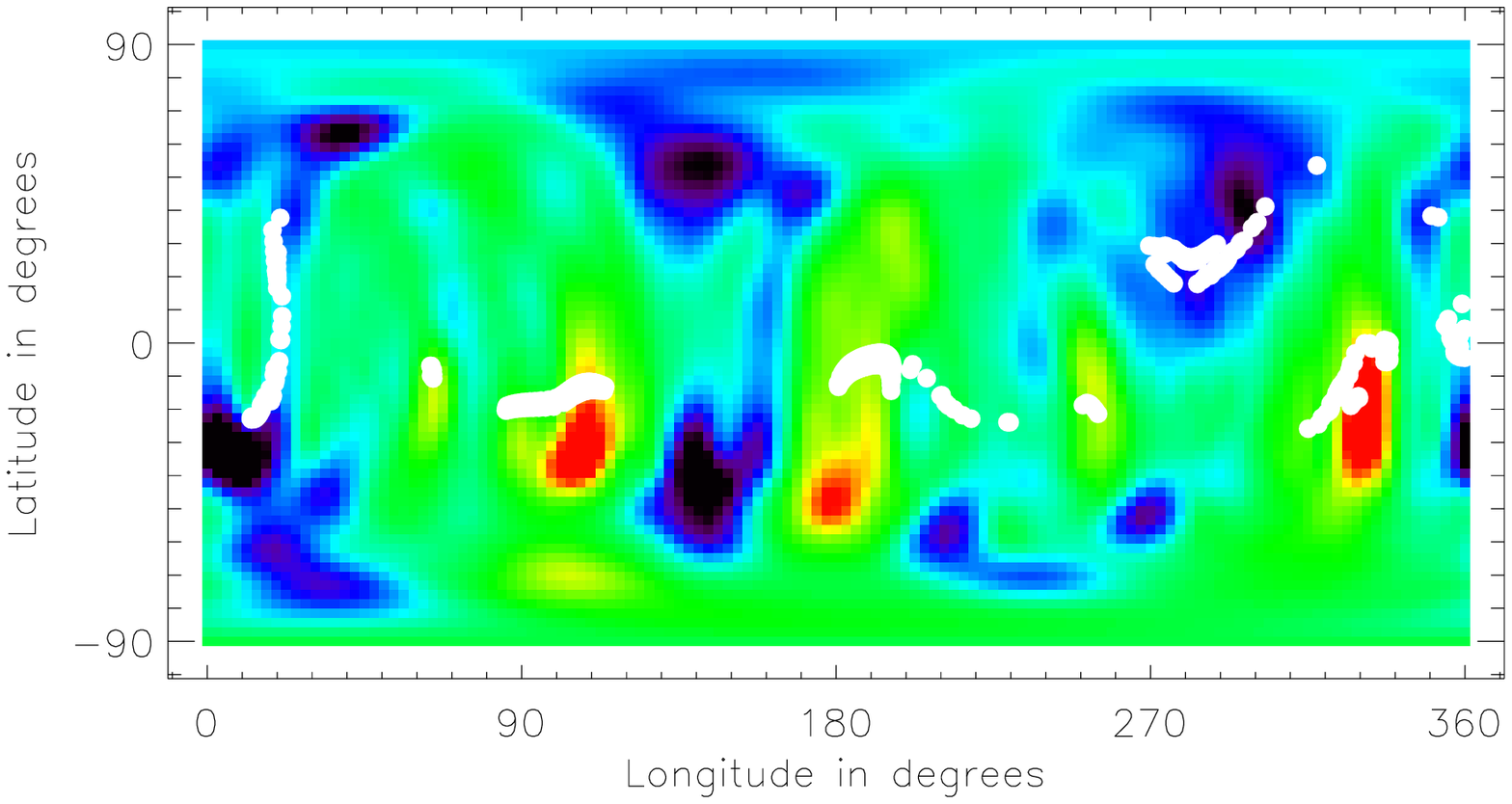,width=85mm}
                        } &
                \subfigure[]{
                        \label{abdor_feet} 
                        \psfig{figure=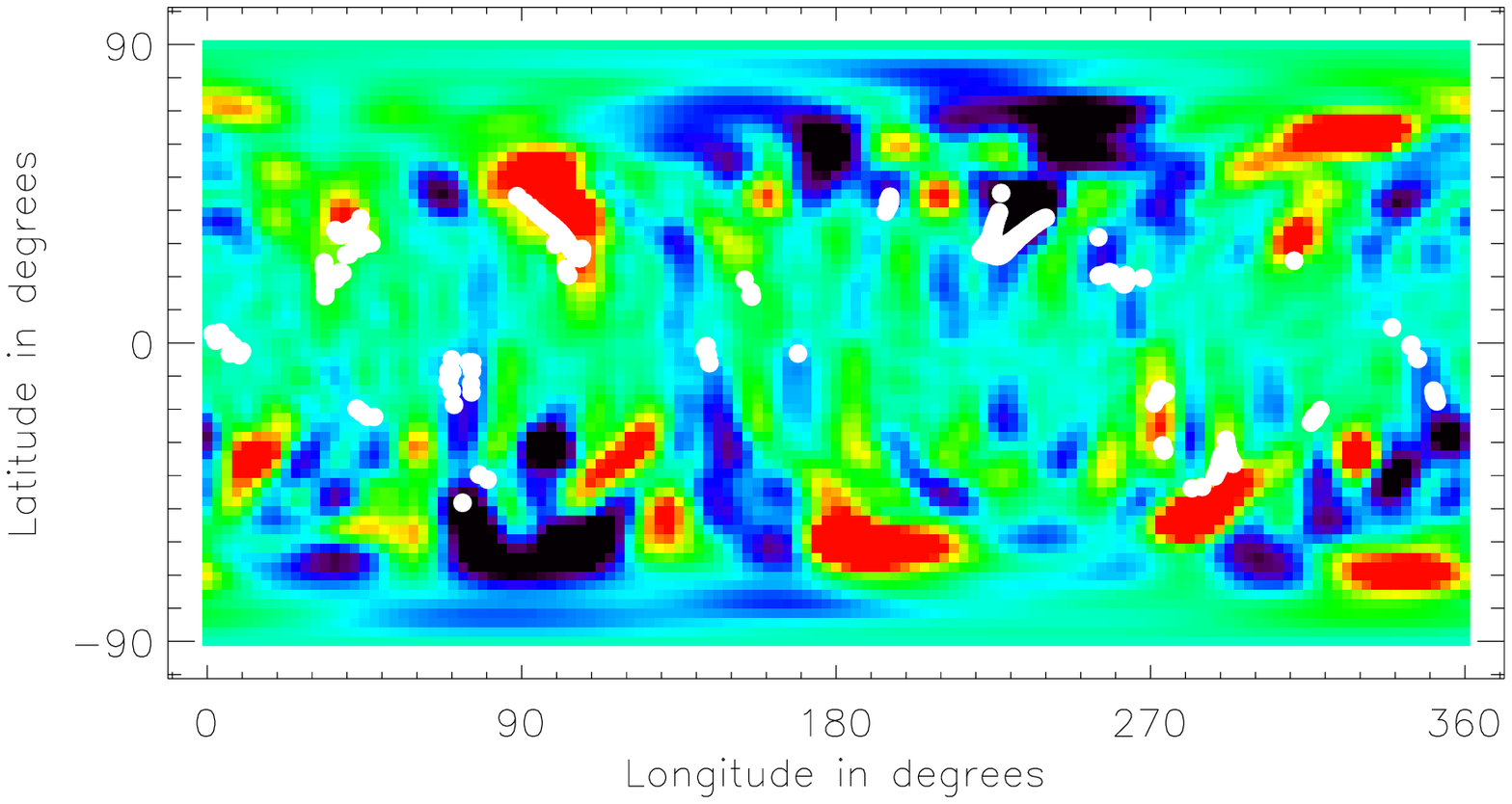,width=85mm}
                        }  \\
        \end{tabular} 
        \caption[]{Surface magnetograms as used in Fig. \ref{fields} for (a) the LQ Hya-like 
                   and (b) the AB Dor-like magnetic fields, coloured to show the strength of the 
                   radial component of the field with red representing 1kG and black -1kG.  White points 
                   are the accreting field line foot points and give an indication of the location 
                   of hot spots.  Hot spots span a range of  
                   latitudes and longitudes; this is in contrast for accretion to a dipole 
                   field where the accreting field line foot points would be at high latitudes 
                   towards the poles.  The accretion filling factor is about 2\% in both 
                   examples.}
    \label{footpoints}
\end{figure*}

The condition for a sonic point to exist on any field line, open or closed, may be
expressed as equation (\ref{sonic}).  $\mathbf{\hat{s}}$ is a unit vector along the
path of the field which may be written as $\mathbf{\hat{s}}=\mathbf{B}/B$, which we 
can use to rewrite (\ref{sonic}) as,
\begin{equation}
c_s^2\frac{dB}{ds}=\mathbf{g}_{eff} \cdot \mathbf{B}.
\label{condition}
\end{equation}
From this it can be seen that the condition for a sonic point to exist depends on 
three things: first, the path that a field line takes through the star's 
gravitational potential well; second, how quickly the strength of the magnetic field
varies as the flow moves along the field line and third, the accretion flow 
temperature which enters through the sound speed, where for isothermal accretion
$c_s^2 \propto T$.  It is the interplay between all three of these factors which 
determines if a sonic point exists.  

For low accretion temperatures the sound speed $c_s$ is small, whilst for open 
field lines the $\mathbf{g}_{eff} \cdot \mathbf{B}$ on the RHS of 
(\ref{condition}) is usually larger than for closed field lines.  This can be seen by
considering the simple example of a closed dipolar field line and a radial open
field line threading the disc at the same point $R_d$.  For both field lines the effective
gravity vector $\mathbf{g}_{eff}$ at $R_d$ is the same.  For the radial open field line the
magnetic field vector $\mathbf{B}$ is aligned with $\mathbf{g}_{eff}$, whereas for the closed
dipolar field line, which threads the disc at a large angle, there is some angle $\theta$ 
between it and $\mathbf{B}$.  Hence the scalar product $\mathbf{g}_{eff} \cdot \mathbf{B} = gB\cos{\theta}$
is larger for the open field line.  Thus for open field lines the RHS of (\ref{condition})
is larger compared to closed field lines, and therefore a higher accretion flow temperature is required
to create the necessary high value of $c_s$ in order to balance the two sides of equation (\ref{condition}).
Of course, for reasons discussed above, $dB/ds$ is larger for the radial field line, but it is not large enough
to compensate for the $\mathbf{g}_{eff} \cdot \mathbf{B}$ term being so small.  At higher values of the accretion
flow temperature more open field lines are able to accrete, thus helping to increase the mass accretion rate 
even though the number of closed accreting field lines is significantly less compared to an 
aligned dipole field.    
             
The aligned dipole field has an accretion filling factor of just under 4\% 
(see Fig. \ref{faccbeta_dipole}), with the $\beta=15\degr$ case having the largest filling factor 
due to the shape of the closed field lines threading the disc, which allows material to be 
channelled onto a larger area of the stellar surface.  More accreting open field 
lines means a larger fraction of the star is covered in accreting field line foot points thus 
increasing the $f_{acc}$ at large accretion flow temperatures.  However, when $\beta$ becomes 
larger as we tilt the dipole further onto its side, $f_{acc}$ decreases.  As the field is 
tilted fewer closed field lines are available for accretion as they no longer intersect the disc 
(see Fig. \ref{closedbeta_dipole}), leading to a decrease in the filling factor 
(see Fig. \ref{faccbeta_dipole}).  The accretion filling factor is smallest for the perpendicular 
dipole where material is accreted onto bars about the star's equator (see Fig. \ref{dipole_feet}), 
compared to the accretion rings about the poles obtained with the aligned dipole.  

We therefore conclude that by considering accretion to tilted dipolar magnetic fields both the mass
accretion accretion rate and the accretion filling factor are dependent on the balance between the 
number of closed and open field lines threading the disc.  If there are many open field lines threading the 
disc, then a higher accretion flow temperature is required in order for the open field to contribute 
to $\dot{M}$.  At high accretion flow temperatures there is at most a factor of two difference between 
the mass accretion rate that the aligned dipole can support compared to the tilted dipoles with large
values of $\beta$.  It appears as though the structure of the magnetic field has only a small role to play 
in determining the mass accretion rate, at least for purely dipolar fields, however, as is discussed in the 
next section, the magnetic field geometry is of crucial importance in controlling the location and 
distribution of hot spots.  

%-------------------------------------------------------------------------------------

\section{Accretion to complex magnetic fields}

\begin{figure*}
        \def\subfigtopskip{4pt}
        \def\subfigbottomskip{4pt}
        \def\subfigcapskip{2pt}
        \centering
        \begin{tabular}{cc}
        \subfigure[]{
                        \label{lqhyahist}                       
                        \psfig{figure=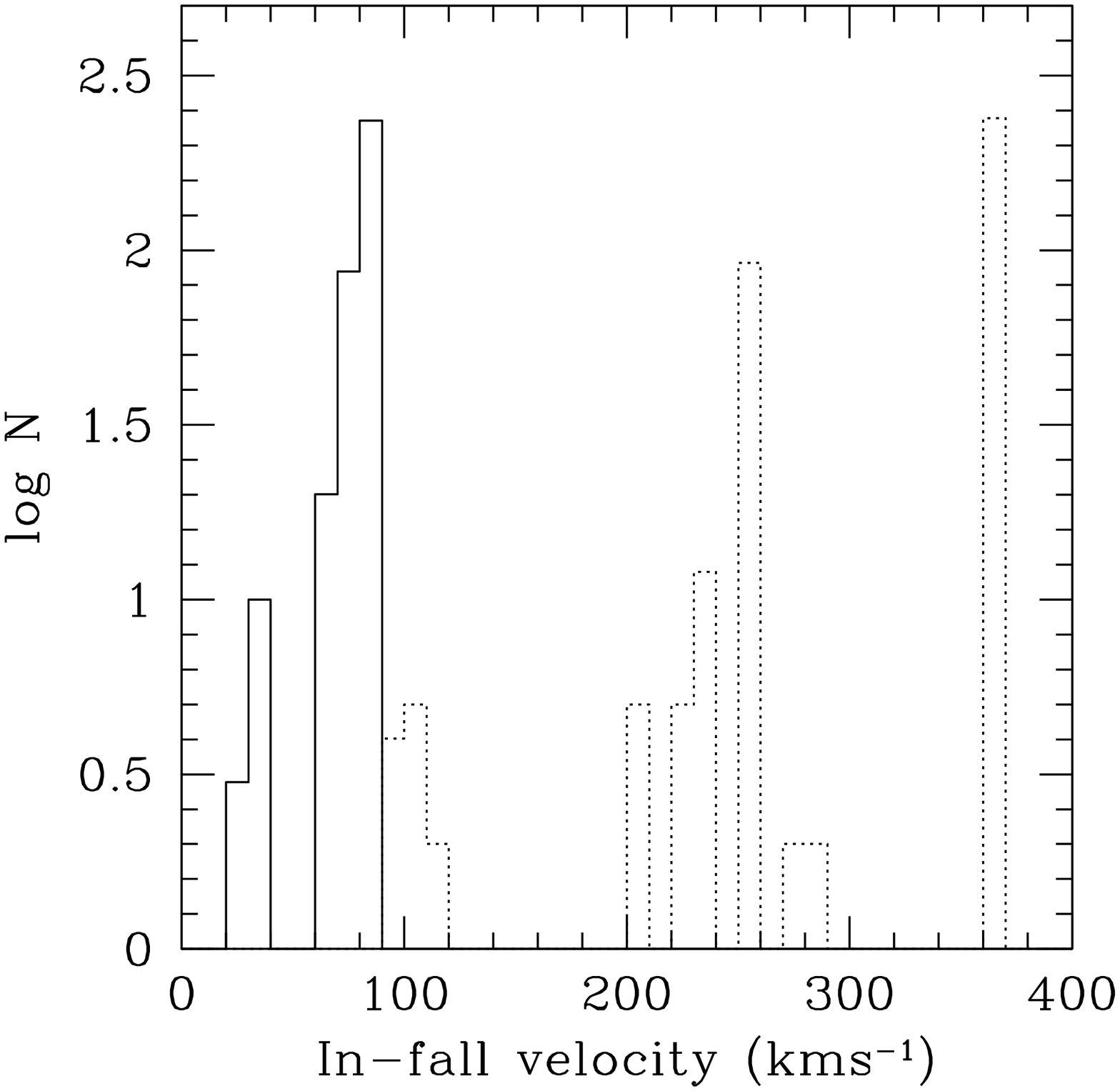,width=80mm}
                        } &
                \subfigure[]{
                        \label{abdorhist} 
                        \psfig{figure=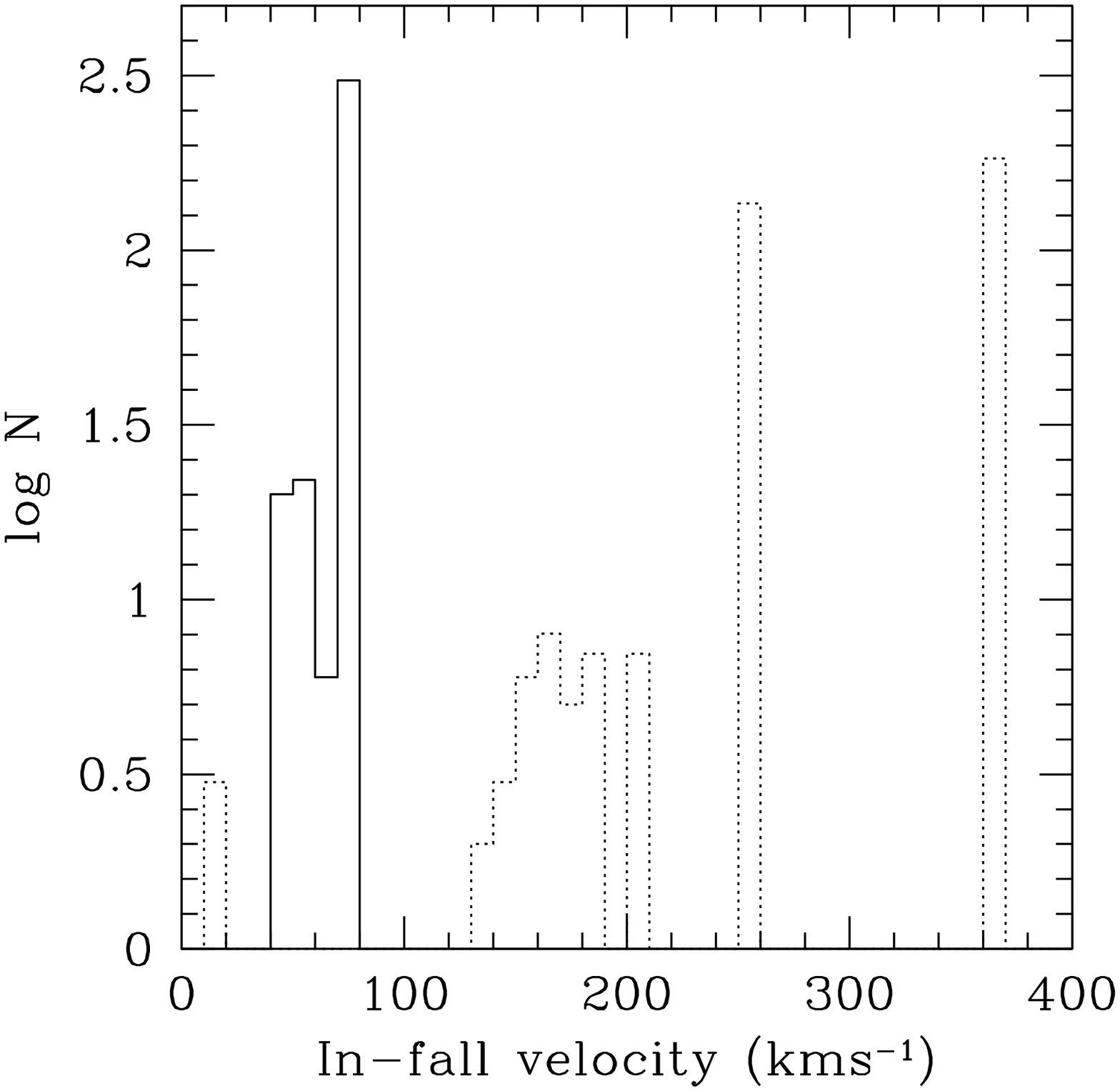,width=80mm}
                        }  \\
        \end{tabular} 
        \caption[]{Histograms showing the distribution of in-fall velocities assuming steady state 
                   accretion flows at 10$^4$K, using the field extrapolations
                   in Fig. \ref{fields}, corresponding to (a) the LQ Hya-like field 
                   and (b) the AB Dor-like field using the DF Tau ({\it solid}) and CY Tau ({\it dots}) 
                   stellar parameters listed in Table \ref{table}.}
    \label{hist}
\end{figure*}

\subsection{Distribution of hot spots}
Brightness modulations have long been interpreted as evidence for hot spots 
on CTTSs (e.g. \citealt{bou95}).  Hot spots are 
a prediction of magnetospheric accretion models and arise from channelled in-falling 
material impacting the star at large velocity.  The distribution of accretion hot spots
and their subsequent effects on photometric variability have been studied by several 
authors, however, only with dipolar magnetic fields (\citealt{woo96}; \citealt{mah98}; 
\citealt{sta99}; \citealt{rom04}).  We find that complex magnetic field 
geometries have a large effect on the location of hot spots.  The  
accreting field line foot points are shown in Fig. \ref{footpoints}, which represent
those field lines in Fig. \ref{fields} which satisfy the accretion conditions discussed 
in \S3.3.  These give an 
indication of how different field geometries control the shape, location and 
distribution of hot spots.  We find a series of discrete hot spots which
span a range of latitudes and longitudes with typical accretion filling factors of
around 2\%.  This is quite different to what we would expect for
accretion to an aligned dipole field, where the accreting field line foot points
would be at high latitudes, towards the poles.  With the complex magnetic fields
presented here hot spots can be at high latitudes, but also often at low latitudes 
close to the star's equator.  The existence of low latitude hot spots has also been 
predicted by \citet{von06} who consider accretion along dynamo generated stellar field 
lines rather than along dipolar field lines.  It is worth noting that the line-of-sight 
field components inferred from polarisation measurements made using the HeI 5876{\AA} emission 
line by \citet{val04}, are well matched by a simple model of a single hot spot at different
latitudes dependent on the particular star.  Such observations already suggest that 
low latitude hot spots may be a common feature of CTTSs.

%-------------------------------------------------------------------------------------

\subsection{In-fall velocities}
In Appendix A we discuss an analytic method for
calculating the location of the critical radius.  This allows us to 
determine which of the field lines from Fig. \ref{fields} have sonic points,
and then find a transonic velocity solution, where material leaves the 
disc at a low subsonic speed but arrives at the stellar surface with a
large supersonic speed.  The accreting field geometry obtained
when considering complex magnetic fields is such that there are many
field lines of different size and shape which are able to support accretion
flows.  This results in a distribution of in-fall speeds, rather than a 
discrete in-fall speed that would be expected for accretion along aligned
dipole field lines.  Fig. \ref{hist} shows the distributions of in-fall
velocities for our two accreting field geometries, and for each set of stellar 
parameters listed in Table \ref{table}, for an accretion flow temperature of 
10$^4$K and a coronal temperature of 10MK.  In both cases the in-fall speeds are 
large enough to produce hot spots on the stellar surface.  We obtain larger
in-fall velocities when using the CY Tau parameters, which is a higher 
mass star, compared to the DF Tau parameters.  Material accreting onto larger mass 
stars will experience a steeper gravitational potential than for accretion onto 
lower mass stars, and this combined with the larger corotation radii for the larger 
mass stars, naturally leads to greater in-fall speeds. 

The accretion flow temperature has a negligible effect on the average in-fall 
velocity, for any magnetic field structure.  We find that the average 
in-fall speed remains almost constant as the accretion flow temperature is varied,
changing by only a few kms$^{-1}$.  As the temperature is 
increased the sonic point of the accretion flow is further from the disc, which
reduces the final Mach number by which material is arriving at the star.  However
this reduction in the final Mach number is caused almost exclusively by the 
increase in the sound speed at higher temperature, whilst the average in-fall velocity
remains constant. 

\begin{figure*}
        \def\subfigtopskip{4pt}
        \def\subfigbottomskip{4pt}
        \def\subfigcapskip{2pt}
        \centering
        \begin{tabular}{cc}
        \subfigure[]{
                        \label{mdot}                         
                        \psfig{figure=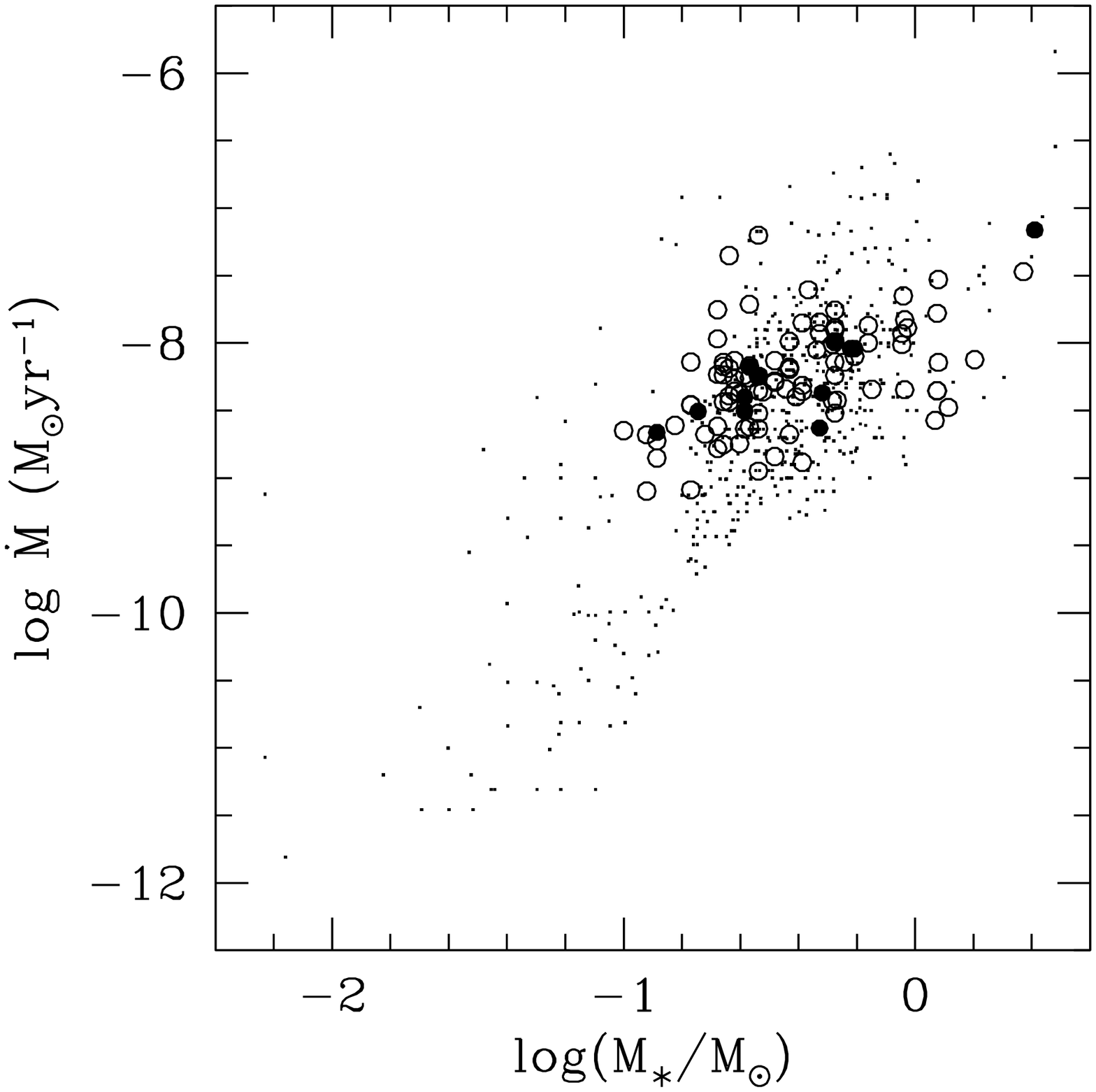,width=80mm}
                        } &
                \subfigure[]{
                        \label{mdotstrong} 
                        \psfig{figure=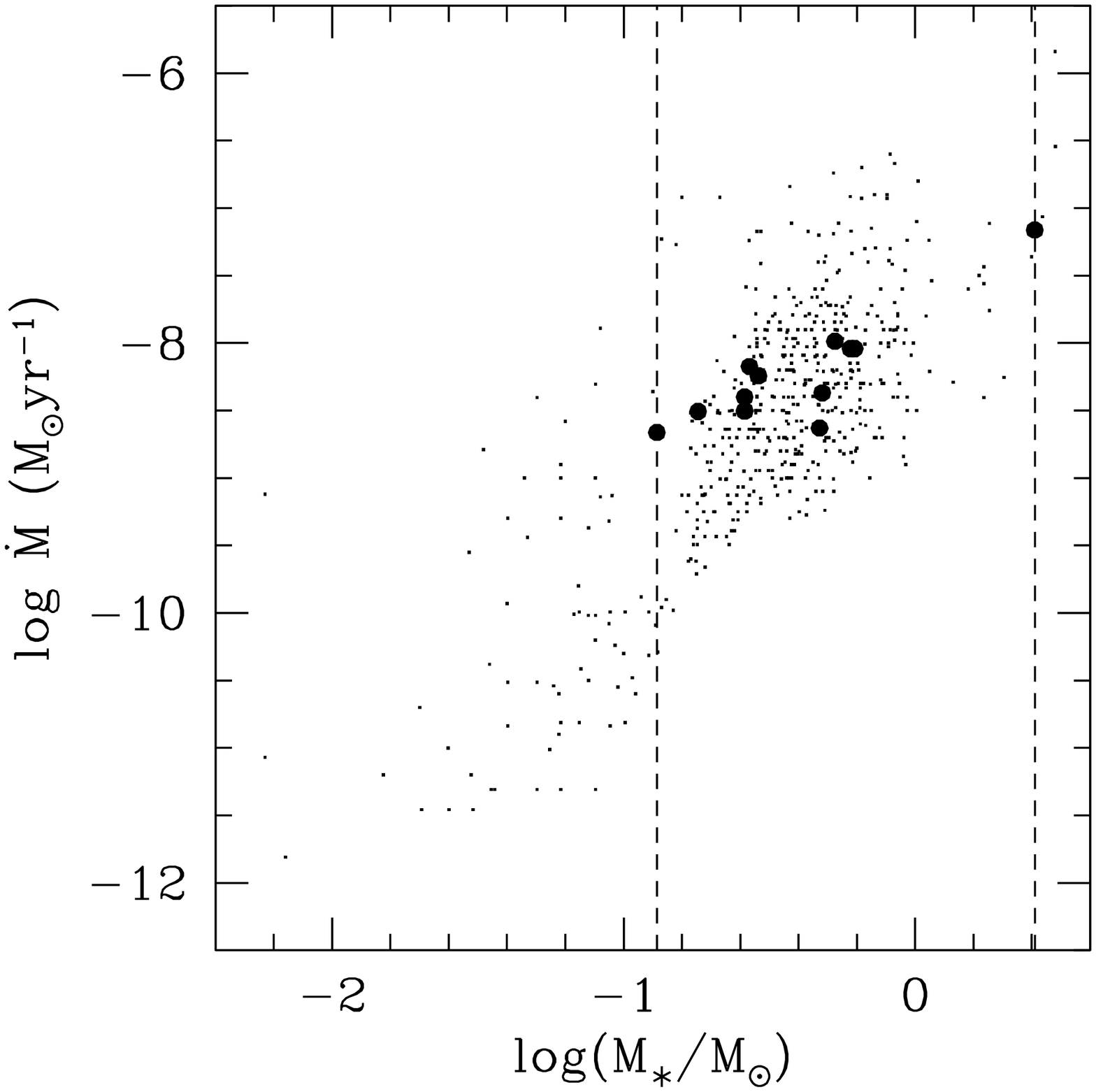,width=80mm}
                        }  \\
        \end{tabular} 
        \caption[]{(a). The correlation between mass accretion rate and stellar mass.  
         Previously published values [{\it points}] are taken from \citet{reb00,reb02}, \citet{moh05}, \citet{nat06}
         and \citet{muz05} which is a collection of data from \citet{gul98}, 
         \citet{whi01}, \citet{whi03}, \citet{muz03}, \citet{cal04} and \citet{nat04}.  
         Using data from the COUP sample of accreting stars \citep{get05} our accretion model
         yields mass accretion rates [{\it large circles}] that are consistent within the observed scatter.  
         {\it Open circles} are values calculated from COUP stars with estimates of $R_{\ast}$, $M_{\ast}$, $P_{rot}$,
         coronal temperature and a CaII equivalent width measurement, whilst the 
         {\it filled circles} are values calculated for stars which are regarded as strongly accreting with
         W(CaII)$<-1$\AA.  All quantities have been calculated using the LQ Hya surface map
         and the higher coronal temperatures derived from the COUP data. (b) shows only
         those stars which are regarded as strongly accreting with the {\it dashed lines} indicating
         the range in mass covered by such COUP stars.}
    \label{mdotmass}
\end{figure*}

There is a broader distribution of in-fall velocities when considering the CY Tau
stellar parameters compared to the narrow distribution for DF Tau, which is more
strongly peaked about a single in-fall velocity (see Fig. \ref{hist}).  We calculate the 
natural radial extent of DF Tau's corona $R_s$ to be larger than the corotation radius
($R_s=2.5R_{\ast}$ for the AB Dor-like field and $2.7R_{\ast}$ for the LQ Hya-like 
field compared to $R_{co}=2.47R_{\ast}$) and therefore accretion proceeds along a mixture 
of open and closed field lines which thread the disc at corotation.  It can be seen from 
Fig. \ref{lqhya_extrap} that accretion occurs almost exclusively from the corotation 
radius with a small range of azimuths where there are neither open nor closed field
lines threading the disc at corotation, and therefore the disc extends closer to 
the star.  This effect is much greater for the CY Tau parameters as its corotation radius
is beyond the natural extent of the corona ($R_s=2.7R_{\ast}$ for the AB Dor-like field and 
$2.9R_{\ast}$ for the LQ Hya-like field compared to $R_{co}=9.55R_{\ast}$).  Most of the
accretion occurs along radial field lines from corotation. However,
at some azimuths there are no open field lines stretching out through the disc at corotation,
and therefore the inner disc at those azimuths is much closer to the star, with accretion
occurring along the closed field lines loops which constitute the star's corona.  A more complete 
model would take account of the torques resulting from 
allowing accretion to occur from well within the corotation radius.  We do not account for this in our model,
however this only occurs for a very small number of field lines, with the bulk of accretion
occuring from corotation.  The number of field lines accreting at lower velocity from within
corotation is small compared to those accreting from $R_{co}$, with the resulting distribution of in-fall 
speeds instead reflecting variations in the size and shape of field lines accreting from 
corotation.        

We have found that the structure of these complex magnetic fields and more importantly the 
stellar parameters (mass, radius, rotation period, coronal temperature) are 
critical in controlling the distribution of in-fall velocities.  Consequently
this will affect mass accretion rates, which we discuss below, and also 
spectral line profiles, which will be the subject of future work.       
   
%------------------------------------------------------------------------------

\subsection{Mass accretion rates and stellar mass}
To investigate the correlation between mass accretion rate and stellar mass we require
a large sample of accreting stars with estimates of stellar mass, radius, rotation period
and coronal temperature.  The X-ray emission from young stars in the Orion Nebula 
has been studied by COUP (The Chandra Orion Ultradeep Project), which provides a vast
amount of data on accreting stars with estimates of all the parameters required
by our model to calculate mass accretion rates.  \citet{get05} provides an overview of the 
observations, and the COUP dataset, which is available 
from ftp://ftp.astro.psu.edu/pub/gkosta/COUP\_PUBLIC/. 

\citet{reb00} first noted that the apparent increase in mass accretion rate
with stellar mass and that the lack of low mass stars with high $\dot{M}$ was a real effect for
stars in the Orion flanking fields.  \citet{whi01} also noted an apparent $\dot{M}-M_{\ast}$ 
correlation for stars in Taurus-Auriga, with a large scatter in $\dot{M}$ values, with
\citet{reb02} reporting that the correlation also existed for stars in NGC 2264.       
The correlation was then found to extend across several orders of magnitude 
in mass with the detection of accretion in low mass T Tauri stars and brown dwarfs 
\citep{whi03} with \citet{muz03} being the first to suggest that $\dot{M} \propto M_{\ast}^2$.
Subsequent observations by \citet{cal04} indicated that this correlation extended to the higher mass, 
intermediate mass T Tauri stars with several authors adding data at lower masses from
various star forming regions (\citealt{nat04}; \citealt{moh05}; \citealt{muz05}), with 
\citet{nat06} adding data from $\rho$-Ophiuchus.  There can be as much as 
three orders of magnitude scatter in the measured mass accretion rate at any particular stellar 
mass.  It should also be noted that mass accretion rate measurements for stars in the Trapezium 
cluster are not consistent with the $\dot{M} \propto M_{\ast}^2$ correlation. 
\citet{rob04} report $\dot{M}$ values for the Trapezium cluster which are significantly lower 
than those obtained for stars in Taurus and the Orion flanking fields and suggest that the probable 
cause is that the discs of lower mass stars are being disrupted by UV radiation from the Trapezium 
OB stars, causing a large drop in mass accretion rates.  Also, \citet{cal04} point out that there 
is a strong bias against the detection of intermediate mass T Tauri stars 
($M_{\ast}=1.5-4M_{\odot}$) with lower mass accretion rates, therefore the exponent of the 
$\dot{M}-M_{\ast}$ correlation may be less that 2.  Further, \citet{cla06} demonstrate that
currently available data are limited by selection effects at high $\dot{M}$ values (whereby 
accretion rates cannot be determined when accretion luminosity is greater than 
stellar luminosity) and also by a lower bound defined by the upper limits of non-detections. 
Their work therefore suggests that the steep correlation between $\dot{M}$ and $M_{\ast}$ is a 
natural consequence of detection/selection limitations, and that the true $\dot{M}-M_{\ast}$
correlation may be different.      

Using the complex magnetic fields discussed above and 
assuming an accretion flow temperature of 10$^{4}$K, we have calculated mass accretion
rates and accretion filling factors for the COUP sample of stars which have estimates 
of $M_{\ast}$, $R_{\ast}$, $P_{rot}$, coronal temperature and measurements of the CaII 8542{\AA} line, 
using the lower coronal temperature, and, for those stars with spectra fitted with a two-temperature model, 
using the higher coronal temperature as well.  We have looked for a correlation between the mass 
accretion rate and stellar mass of the form $\dot{M} \propto M_{\ast}^{\alpha}$, where 
$\alpha$ is a constant. In Fig. \ref{mdot} we have plotted
our calculated mass accretion rates from the COUP stars as a function of stellar mass, 
over-plotted on published values, for the LQ Hya-like magnetic field using the higher coronal
temperatures.  \citet{pre05} describe 
how the equivalent width of the CaII 8542{\AA} line is used as an indicator of accretion for stars in the 
COUP dataset.  They follow the classification discussed by \citet{fla03}, who assume that stars are 
strongly accreting if the CaII line is seen in emission with W(CaII)$<-$1{\AA}.  Stars with
W(CaII)$>$1{\AA} are assumed to be either weak or non-accretors.  However, it should be noted that
many of the stars in the COUP dataset have W(CaII)$=$0{\AA}, and as such cannot be identified
as either accreting or non-accreting.  \citet{sta06} discuss, with particular reference to the
COUP, how some stars can show clear accretion signatures in H$\alpha$ but without showing evidence 
for accretion in CaII.  Therefore the sample of COUP stars
considered in Fig. \ref{mdot} may not be restricted to actively accreting CTTSs and could also include other 
non-accreting young stars, such as weak line T Tauri stars, where the disc is rarefied or non-existent, 
and stars which are surrounded by discs but are not actively accreting at this time.  However, we
consider all of the available stellar parameters from the COUP dataset in order to demonstrate that
our model produces a similar amount of scatter in calculated $\dot{M}$ values.  In Fig. \ref{mdotstrong} 
we have only plotted those COUP stars which are regarded as strongly accreting based on the equivalent 
width of the CaII 8542{\AA} line.
   
By fitting a line to the filled circles in Fig. \ref{mdotstrong} we find a correlation of the 
form $\dot{M} \propto M_{\ast}^{1.1}$ for COUP stars which are regarded as strong accretors,
using the LQ Hya-like magnetic field.  For the AB Dor-like field we find 
$\dot{M} \propto M_{\ast}^{1.2}$.  Therefore our simple steady state isothermal accretion model 
produces an increase in mass accretion rate with stellar mass, and predicts $\dot{M}$ values which 
are consistent within the observed scatter, but it underestimates the exponent of 2 obtained from
published values.  We find similar results when considering both the lower and higher coronal temperatures. 
However, the strongly accreting COUP stars only include
stellar masses of $M_{\ast}\approx 0.1-2.6M_{\odot}$, and as such do not provide
a large enough range in mass to test if our accretion model can reproduce the observed
correlation.  Therefore if we only consider the observational data in the restricted range of mass 
provided by the COUP dataset (indicated by the vertical dashed lines in Fig. \ref{mdotstrong}) then 
the exponent of the observed correlation is less than 2, with $\dot{M} \propto M_{\ast}^{1.4}$, and if 
higher mass stars do exist with lower mass accretion rates, could even be less than 1.4.  Therefore our 
simple model, which \citet{jar06} have successfully used to explain the observed correlation between 
X-ray emission measure and stellar mass, comes close to reproducing the stellar mass - 
mass accretion rate correlation.  It is also worth noting that our model produces a correlation in 
agreement \citet{cla06}.       

\begin{figure}
\centering
\psfig{file=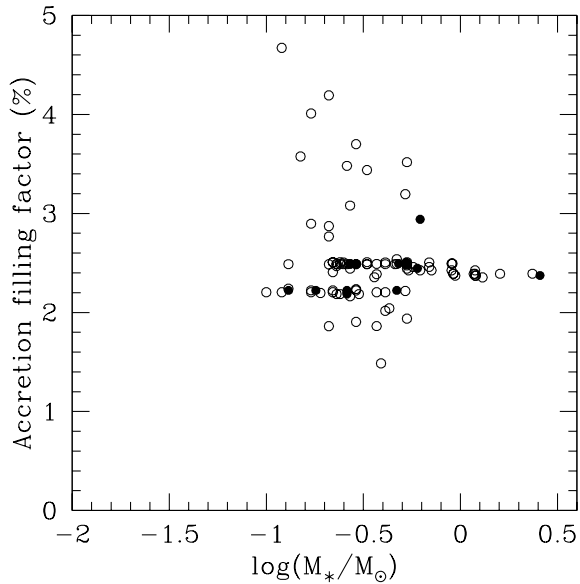,width=80mm}
\caption{The accretion filling factors are small, and typically
         around $2.5$\% for the LQ Hya-like magnetic field.  {\it Open} and {\it filled circles} 
         represent the same set of stars in Fig. \ref{mdotmass}.}
\label{faccmass}
\end{figure}

The accretion filling factors are typically around 2.5\% for the sample of
COUP stars (see Fig. \ref{faccmass}), although it varies from less than 2\% to greater 
than 4\%.  There is a slight trend for higher mass stars to have smaller filling 
factors, that is there is a slight trend for stars with larger corotation
radii to have a smaller filling factors despite their larger mass accretion rates.  
For stars with smaller corotation radii (lower mass) there are many field lines 
threading the disc and able to support accretion, and therefore a large fraction of the 
stellar surface is covered in accreting field lines foot points, whereas for stars with 
larger corotation radii (higher mass) there are less accreting field lines and therefore
smaller filling factors.  However the higher mass stars are accreting at a larger velocity
therefore producing larger mass accretion rates.  The actual accretion filling factor depends
on the magnetic field structure.  For the AB Dor-like magnetic field the accretion filling factors
are smaller at around 1.8\%.  

%-----------------------------------------------------------------------------------

\section{Summary}
By considering accretion to both dipolar and complex magnetic fields we have constructed a
steady state isothermal accretion model were material leaves the disc
at low subsonic speeds, but arrives at the star at large supersonic speeds.  We find
in-fall velocities of a few hundred kms$^{-1}$ are possible which is consistent with 
measurements of the redshifted absorption components of inverse P-Cygni profiles 
(e.g. \citealt{edw94}).  We found that for accretion along aligned and perpendicular
closed dipole field lines that there was little difference in in-fall speeds between the
two cases.  However, for tilted dipole fields in general (rather than just considering 
a single closed field line loop) there are many open and closed field lines threading the 
disc providing different paths that material could flow along from the disc to the star.
The path that a particular field line takes through the star's gravitational potential
combined with the strength of magnetic field and how that changes along the field line
path, are all important in determining whether or not a field line can support transonic
accretion.  At low accretion flow temperatures open field lines are typically not able to 
support transonic accretion, but at higher accretion flow temperatures they can accrete 
transonically and consequently we see an increase in the mass accretion rate.
We only find a factor of two to three difference in $\dot{M}$ values for accretion
to tilted dipolar magnetic fields suggesting that the geometry of the field itself
is not as significant as the stellar parameters (which control the position of corotation)
in controlling the mass accretion rate.  

The magnetic field geometry is crucial in controlling the location and distribution
of hot spots on the stellar surface.  For accretion to complex magnetic fields
we find that hot spots can span a range of latitudes and longitudes, and are
often at low latitudes towards the star's equator.  We find that the accretion
filling factors (the fractional surface area of a star covered in hot spots) 
are small and typically around 2.5\%, but they can 
vary from less than 1\% to over 4\% and rarely to larger values.  This is consistent
with observations which suggest small accretion filling factors and 
the inference of hot spots at various latitudes \citep{val04}.

For accretion with complex magnetic fields there is a distribution of in-fall speeds,
which arises from variations in the size and shape of accreting field lines.  The resulting 
effect that this will have on line profiles will be addressed in future work.
In our model most of the accretion occurs from the corotation radius, but at some azimuths the
disc extends closer to the star meaning that a small fraction of field lines are accreting
material at lower velocity.  Lower mass stars, with their lower surface gravities, typically have 
larger coronae which would extend out to corotation \citep{jar06}.  Therefore the lower mass 
stars are accreting along a mixture of open and closed field lines from the corotation radius.  
In contrast higher mass stars, with their higher surface gravities, have small compact coronae, 
and so the star is actively accreting along mainly open field lines from the corotation radius.  However,
such open field lines do not thread the disc at all azimuths, with some accretion instead occuring along
field lines which are much closer to the stellar surface.  This gives rise to a 
small peak at low velocity in the in-fall velocity distribution.  However, this represents 
only a small fraction of all accreting field lines which do not contribute significantly to the 
resulting mass accretion rate.

Finally we applied our accretion model to stars from the COUP dataset which have estimates of 
the stellar parameters and measurements of the equivalent width of the CaII 8542{\AA} line, 
which is seen in emission for accreting stars.  For the complex magnetic fields we calculated
mass accretion rates and accretion filling factors as a function of stellar mass.  The observed
stellar mass - accretion rate correlation is $\dot{M}\propto M_{\ast}^2$ 
(\citealt{muz03}; \citealt{cal04}; \citealt{moh05}; \citealt{muz05}), however, this may be 
strongly influenced by detection/selection effects \citep{cla06}. By only considering
observational data across the range in mass provided by the COUP sample of accreting stars, the
observed correlation becomes $\dot{M}\propto M_{\ast}^{1.4}$.  Our steady state isothermal model
gives an exponent of 1.1 for the LQ Hya-like magnetic field and 1.2 for the AB Dor-like field with 
similar results for both the high and low coronal temperatures, with the caveat that the observed
correlation may be less than 1.4 due to a strong bias against the detection of higher mass
stars with lower mass accretion rates \citep{cal04}.  It may be the case that an exponent of 1.2 
compared to the observed 1.4,
represents the best value that can be achieved with a steady state isothermal accretion
model.  \citet{jar06} have used this model to reproduce the observed increase in X-ray emission 
measure with stellar mass \citep{pre05}. However, they find that when using the complex magnetic fields presented
here (extrapolated from surface magnetograms of the young main-sequence stars AB Dor and LQ Hya) 
they slightly under estimate the emission measure-mass correlation.  When they use dipolar 
magnetic fields, which represent the most extended stellar field, they slightly over estimated the 
correlation.  This suggests that T Tauri stars have magnetic fields which are more extended
than those of young main sequence stars, but are more compact than purely dipolar fields.  

Also, our model has not been tested across a large enough range in mass to make the comparison with 
the observed correlation of $\dot{M} \propto M_{\ast}^2$.  However, it already compares well
with the alternative suggestion of \cite{cla06} that the $\dot{M}-M_{\ast}$ correlation
is not as steep with $\dot{M} \propto M_{\ast}^{1.35}$.  In future we will extend our model to
consider accretion from the lowest mass brown dwarfs, up to intermediate mass T Tauri stars.  
This will require estimates of, in particular, rotation periods and coronal temperatures.          

Our accretion model reproduces a similar amount of scatter in calculated $\dot{M}$ values compared
with observations.  This can be attributed to different sets stellar parameters changing the
structure of the accreting field.  The stellar parameters control the location of the corotation radius,
whilst it is the position of corotation relative to the natural coronal extent, which determines 
whether or not accretion occurs predominantly along the open field.   However, for the moment we are 
restricted to using surface magnetograms of young main sequence stars which may not necessarily represent
the true fields of T Tauri stars.  It could be possible that variations in magnetic field geometry 
from star to star are responsible for the observed large scatter in the mass accretion rate at any 
particular stellar mass.  If there is a large difference in the structure of the accreting field from 
star to star this could lead to large variations in mass accretion rates onto different stars.  The 
spectropolarimeter at the Canada France Hawaii telescope, 
ESPaDOnS: Echelle SpectroPolarimetric Device for the Observation of Stars
(\citealt{pet03}; \citealt{don04}), will in the near future allow the reconstruction of the
magnetic field topology of CTTSs from Zeeman-Doppler imaging.  This will allow a number of
open questions to be addressed about the nature of the magnetic fields of T Tauri stars and
allow us to test if our accretion model can indeed reproduce the stellar mass - accretion
rate correlation.  

%----------------------------------------------------------------------------------

\section*{Acknowledgements}
The authors thank the referee, Chris Johns-Krull, for constructive comments which
have improved the clarity of this paper, and Aad van Ballegooijen who wrote the 
original version of the potential field extrapolation code.  SGG acknowledges the 
support from a PPARC studentship.
\bibliographystyle{mn2e}
\bibliography{accretion}

%----------------------------------------------------------------------------------

\appendix

\section[]{Algorithm for finding a sonic point and the initial Mach number 
           required for a smooth transonic solution}

In order to determine the initial Mach number which results in a
transonic accretion flow we need to understand exactly what the
different solutions in Fig. \ref{velofig} represent.  These curves are
obtained from the pressure solutions plotted in Fig. \ref{prefig} using equation (\ref{velo}).
These pressure profiles track how the roots of (\ref{pre}) 
change as the flow moves along a field line.  The pressure function (\ref{pre}) can be 
written in the form,
\begin{equation}
f \left(\frac{p}{p_d}\right) = \ln{\left( \frac{p}{p_d}\right)}+ 
         a\left( \frac{p_d}{p} \right )^2+b=0,
\label{function}
\end{equation}
where $a$ and $b$ are constants at any fixed point along a field line with,
\begin{eqnarray}
a &=& \frac{1}{2}{\cal M}^2 \left (\frac{B}{B_d}\right )^2, \\
b &=& -\frac{1}{2}{\cal M}^2 - \frac{1}{c_s^2}\int \mathbf{g}_{eff} \cdot \mathbf{\hat{s}}ds.
\end{eqnarray} 

For a Mach number prescribed to a flow leaving the disc, the pressure 
function has two roots, which consequently yield two velocities.  One of the 
roots represents the true physical solution, and the other is a mathematical
solution of no physical significance.  The branches in
Fig. \ref{velofig} occur in pairs, with an individual branch
tracking how one of the pressure roots changes as the flow travels
along the field line.  For example, suppose we
had chosen an initial (subsonic) Mach number which resulted in a 
purely subsonic flow from the disc to the star (curve A in Fig. \ref{velofig}).  
Then one of the roots, the one which gives a subsonic solution everywhere along the flow, is
the true physical one.  The second root of (A1) produces a supersonic 
solution (curve B), which is of mathematical interest, but has no physical meaning.
Conversely, had we chosen the initial Mach number which corresponded
to the same supersonic branch, then this purely supersonic branch is 
the one with physical meaning, with the other root then generating the
purely subsonic branch, the mathematical artifact.  Thus the purely subsonic
and supersonic solutions exist provided our pressure function
(\ref{function}) has two real roots when evaluated at each point
along a field line.

At small values of $p/p_d$ the second $(p/p_d)^{-2}$
term in (\ref{function}) dominates with $f(p/p_d) \to \infty$ as $p/p_d \to 0$.  
At larger values of $p/p_d$ the logarithmic term dominates.  Thus
(\ref{function}) has a minimum which occurs when,
\begin{equation}
\frac{p}{p_d} = {\cal M} \frac{B}{B_d}. 
\label{minimum}
\end{equation}
For this value of $p/p_d$ the pressure function reduces to, 
\begin{equation}
f_c(r) = \ln{\left ( {\cal M} \frac{B}{B_d}\right )} + \frac{1}{2}
-\frac{1}{2} {\cal M}^2
         -\frac{1}{c_s^2} \int \mathbf{g}_{eff} \cdot \mathbf{\hat{s}} ds,
\label{crit}
\end{equation}
which we will refer to as the critical function, $f_c$. In order for two real roots 
to exist at some point $r$ along a field line then $f_c(r)<0$ must hold. Thus for purely 
subsonic$/$supersonic solutions to exist then $f_c<0$ at each point along a
field line, with $f_c$ reaching its highest value at the critical
radius (where the two roots of the pressure function, and
therefore the velocity roots, are closest together).  Another pair
of solutions are the discontinuous ones labelled C and D in Fig.
\ref{velofig}.  For these solutions $f_c<0$ for $r_1 <r\leq R_d$ and
$R_{\ast} \leq r < r_2$.  However for the domain $r_2 < r < r_1$,
$f_c > 0$, and therefore the pressure function has no real roots,
and we cannot satisfy (\ref{pre}); hence there are no velocity
solutions for these values of $r$.  For $r=r_1$ and $r=r_2$, $f_c =
0$, and at these points the pressure function has a single repeated root that
coincides with the pressure function minimum.

\begin{figure}
\centering
\psfig{file=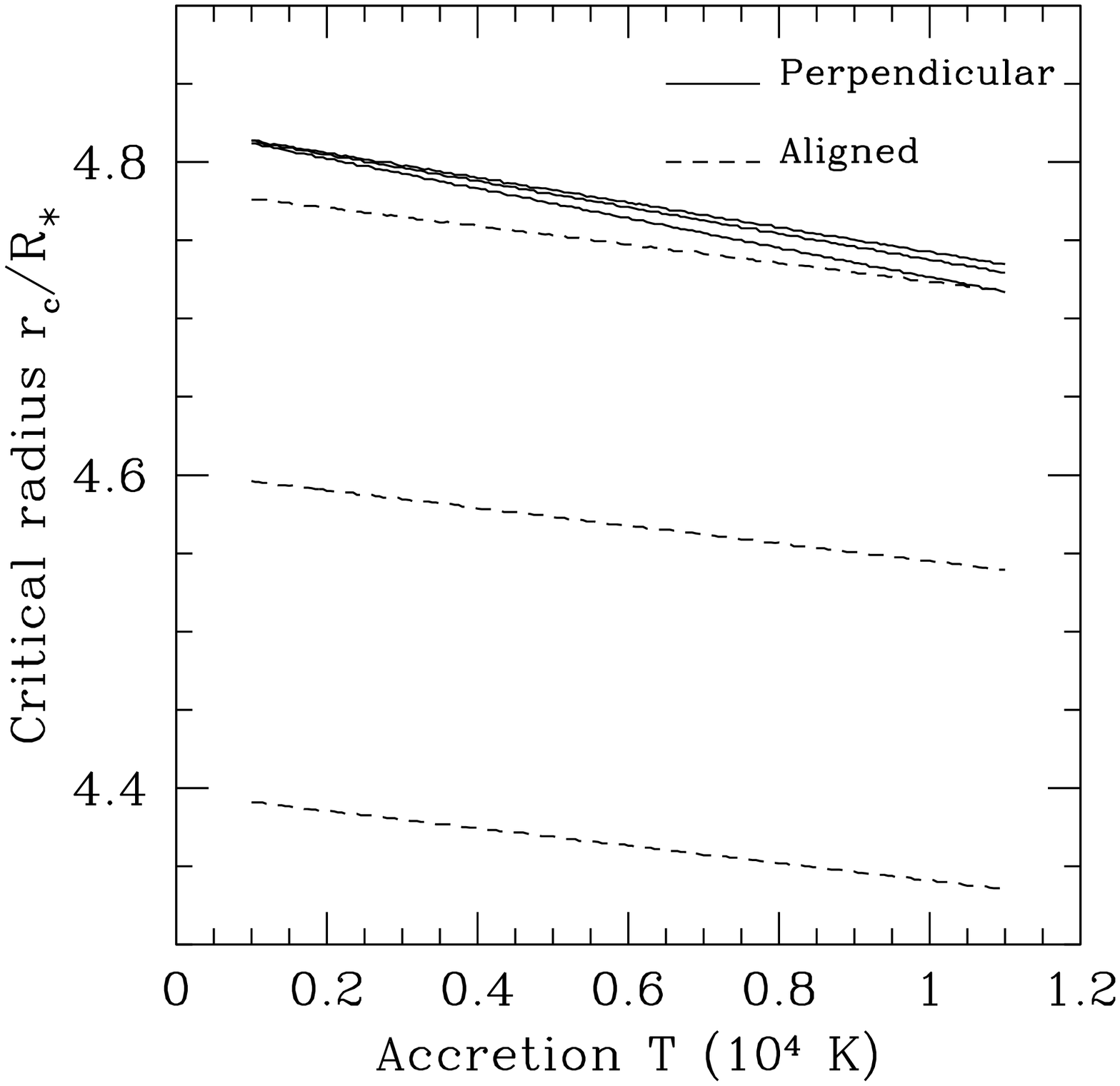,width=80mm}
\caption{The location of the critical radius $r_c$ as a function of accretion flow temperature
        $T_{acc}$ for starting radii $R_d$ about the corotation radius (upper line, 
        $R_d = 7.0R_{\ast}$, middle line $R_d =6.0R_{\ast} \approx R_{co}$ and lower
        line $R_d = 5.0R_{\ast}$).  At lower temperatures $r_c$ is closer to the inner 
        edge of the disc, and the difference between the aligned and perpendicular dipoles 
        is larger.  Variation of the starting radius has little effect for the 
        perpendicular dipole, and is of minor significance for the aligned dipole.}
\label{critfig}
\end{figure}

The final pair of solutions are the transonic ones labelled E and
F in Fig. \ref{velofig}.  Here $f_c < 0$ for all $r$, except at the sonic
point where $r=r_c$ and $f_c = 0$.  Hence, for a transonic
solution the maximum value of the critical function $f_c$ is zero
at the critical radius $r_c$, but less than zero at all other points along
the field line.  This then gives us a robust method for finding
both the critical radius and the initial Mach number required for
a smooth transonic solution (provided such a solution exists).

At the critical radius the minimum of our pressure function is at
its highest value; that is $f_c$ has a maximum turning point.
Hence to determine if a transonic solution exists we select any initial
Mach number ${\cal M}$ and calculate how $f_c$ varies as we move along
a field line from the disc to the star.  If a sonic point exists
on that field line, $f_c$ will have a distinct maximum at some
point.  The advantage of this method is that it is
possible to recover the initial Mach number that would result in a
smooth transonic flow, simply by varying ${\cal M}$ until $f_c = 0$ at
$r_c$; that is we change ${\cal M}$ until the maximum value of $f_c$
occurs at zero. This algorithm is an efficient method for quickly
determining both the critical radius and the initial Mach number which
results in a transonic accretion flow. Of course, not all field lines have a sonic
point (see \S3.3) as the critical radius may either be
interior to the star or exterior to the starting radius.  This algorithm
may be applied to accretion flows along field lines of any size, shape and 
inclination, even in the absence of analytic descriptions of the magnetic
field and effective gravity.

The location of the critical radius (or the sonic point for a
transonic solution) determines the velocity with
which material impacts the stellar photosphere. We have found the critical radius from the 
maximum value of $f_c$ for both 
the perpendicular and aligned dipoles for a CTTS with a mass of 0.5M$_{\odot}$, 
radius 2R$_{\odot}$ and a rotation period of 7 days.  For our two dipole cases 
the $B/B_d$ and $\mathbf{g}_{eff} \cdot \mathbf{\hat{s}}$ terms that contribute to 
(\ref{crit}), have analytic forms given respectively by (\ref{bfield}) and either 
(\ref{grav_equ}) for the perpendicular dipole, or (\ref{grav_perp}) for the aligned 
dipole; in both cases $\mathbf{\hat{s}}=\mathbf{B}/B$.  The location of $r_c$ therefore
changes as we vary both the starting radius $R_d$ and also the temperature (which enters
through the sound speed).  It should be noted, however, that for certain choices
of parameters sometimes $f_c$ has no distinct maximum turning point, indicating that 
either the critical radius is interior to the star ($r_c \leq R_{\ast}$), or 
beyond the starting radius ($r_c > R_d$). In these cases flows that leave the
disc at a subsonic speed remain subsonic all the way to the star,
and likewise supersonic flows remain supersonic at all points
along a field line.

Fig. \ref{critfig} shows how the critical radius varies for a range of
temperatures and starting radii for both the perpendicular and
aligned dipoles.  In both cases, the critical radius moves towards the
inner edge of the disc with decreasing temperature.  In other
words as the accretion flow temperature decreases the sonic point moves along the
field line away from the star, closer to where the flow leaves the
disc.  It is straight forward to explain why this should happen by
considering a transonic flow which would leave the disc at a
subsonic speed.  As the temperature drops the sound speed
decreases ($c_s \propto \sqrt{T}$); therefore a flow leaving the
disc does not have to accelerate for as long to reach its own
sound speed and becomes supersonic sooner.  Thus $r_c$ moves
closer to $R_d$ with decreasing temperature; and for a high enough
temperature the critical radius is interior to the star, and
conversely for a low enough temperature the critical radius is
beyond the starting radius.  However, at least for this particular
set of stellar parameters, the actual change in critical radius location
with temperature is small.    

We also found that as the critical radius moved towards the
inner edge of the disc, the Mach number with which the flow arrived at the
star increased. This would be expected however as the
critical radius moves away from the star due to the sound speed
decrease, which would naturally increase the flow Mach number at
the star (for transonic and purely supersonic flows). 

It can be seen from Fig. \ref{critfig} that for the
perpendicular dipole varying the starting radius of the flow, $R_d$,
has little effect on the location of the critical radius.
Therefore changing the size of equatorial field lines has a
negligible effect on the velocity with which accretion flows
(along the field) reach the star.  For the aligned dipole 
changing where the field lines thread the disc does have an effect 
on the critical radius location and therefore on the final velocity 
with which material impacts the star. However, due to the overall small 
change in the critical radius location evident in Fig. \ref{critfig} 
the in-fall velocity only varies by around 6 kms$^{-1}$. It therefore appears 
that the physical size of field lines, in a dipole accretion model, is 
of little importance for poloidal field lines (aligned north-south in the 
star's meridional plane), and is negligible for equatorial field lines 
(aligned east-west in the star's equatorial plane).

The effect of the field orientation is also of little importance for closed dipolar field
lines.  If we have a closed field line in the equatorial plane, with a maximum radial
extent $R_d$, then changing the inclination of that field line so
that it now lies in the meridional plane, has an effect on
the critical radius location (see Fig. \ref{critfig}). The only major
difference between the two cases enters through the effective
gravity, which only has a radial component in the equatorial
plane, but both $r$ and $\theta$ components in the meridional
plane.  However, the difference in in-fall velocities is again very small, 
at only a few kms$^{-1}$.  This suggests that for accretion along closed dipolar field lines, 
where material is leaving the disc at a fixed radius, the field geometry has little 
effect on in-fall velocities; however, as discussed in \S4.2, this result
does not hold when we consider more complicated multipolar fields.

\bsp

\label{lastpage}

\end{document}